# Pressure dependence of olivine grain growth at upper mantle conditions

Filippe Ferreira[1*+], Marcel Thielmann[1], Robert Farla[2], Sanae Koizumi[3], Katharina Tinka Marquardt[4*]

[1]Bayerisches Geoinstitut, Universität Bayreuth, Bayreuth, Germany, [2]Deutsches Elektronen-Synchrotron (DESY), Hamburg, Germany [3]Earthquake Research Institute, Tokyo, Japan, [4]Department of Materials. University of Oxford, Oxford, UK [+]Now at Bayerisches Landesamt für Umwelt, Augsburg, Germany *katharina.marquardt@materials.ox.ac.uk

Key points

- Grain growth rate in olivine and pyroxene aggregates decreases as pressure increases.
- Reduced grain growth rates at high pressure may be counteracted by the effect of increasing temperature, which accelerates grain growth.
- Change to grain-size sensitive deformation might occur at shallower depths than previously expected.

## Arxiv abstract:

The grain size of olivine influences mass and heat flux in Earth's upper mantle. We performed annealing experiments on synthetic olivine-pyroxene aggregates (6-13 vol.%) at 1-12 GPa and 1323-1793 K. Grain-size analysis via EBSD reveals an activation volume of $4.8 \times 10^{-6}\ \mathrm{m}^3/\mathrm{mol}$, matching silicon grain-boundary diffusion values. This suggests pressure-driven reduction in olivine growth rates may offset temperature effects at depth, enabling grain-size-sensitive creep at shallower mantle depths than previously modeled.

## Abstract

The grain size of olivine influences several processes in the Earth's upper mantle, such as mass and heat flux. However, grain growth, one of the main processes controlling grain size, is still poorly constrained for olivine at upper mantle conditions. To evaluate the effects of pressure on grain growth of olivine, we performed annealing experiments using hot-isostatically-pressed synthetic aggregates of olivine plus 6 and 13 vol.% of pyroxene. The experiments were performed at pressures ranging from 1 to 12 GPa and temperatures from 1323 to 1793 K, using a piston-cylinder and multi-anvil apparatus. We determined grain-size distributions for all experimental run products using Electron backscatter diffraction orientation mapping. The best fit to the resulting data requires an activation volume of 4.8x10$^{-6}$ m$^3$/mol. This value is similar to previously reported activation volumes for silicon grain-boundary diffusion at high pressures. This indicates that silicon grain-boundary diffusion likely controls dry, melt-free olivine grain growth in the upper mantle. Our data show that the olivine grain growth rate in olivine pyroxene aggregates is reduced as pressure increases. These results suggest that with increasing depths in the Earth's upper mantle, reducing grain growth rates due to increasing pressure may offset the temperature effect that should cause grains to grow faster with increasing temperature. Consequently, this may result in smaller average grain sizes than previously anticipated. Because grains remain smaller across deeper parts of the mantle, grain-size-sensitive diffusion creep can occur at shallower depths than previously expected.

## Plain language summary

The grain size of olivine, the primary mineral phase in the Earth's upper mantle, influences several processes in this region, such as mass and heat flux. However, we lack information on how pressure affects grain growth, one of the main processes controlling grain size at depths between 200 and 400 km. To evaluate the effects of pressure on the grain growth of olivine, we performed grain growth experiments using synthetic aggregates of olivine and pyroxene. The experiments were performed at pressures ranging from 1 to 12 GPa and temperatures from 1323 to 1793 K, using a high-pressure apparatus. After the experiments, we determined the grain-size distributions using electron microscopy. Our data demonstrate that the olivine grain growth rate is reduced as pressure increases. With increasing depths in the Earth's upper mantle, the increase in temperature leads to increasing grain growth rates. However, our results suggest that reducing the grain growth rate of olivine due to increasing pressure may offset the effect of temperature on grain growth. Consequently, smaller average grain sizes could be maintained in the middle to deep upper mantle, promoting deformation mechanisms dependent on grain size.

## Keywords

Olivine; grain growth; grain size; high pressure; upper mantle; viscosity.

## Introduction

Grain size is one of the main factors affecting rock viscosity and therefore plays an important role in different geodynamic processes. Through its influence on rock viscosity, grain size impacts the Earth's heat flux and thus its thermal evolution (Hall and Parmentier, 2003; Rozel, 2012; Solomatov, 2001), the Earth's chemical mixing and the formation of heterogeneities in the Earth's mantle (Solomatov and Reese, 2008), the dynamics of subduction slabs and plumes (Dannberg et al., 2017) and localization of deformation (Mulyukova and Bercovici, 2019; Thielmann, 2018; Thielmann et al., 2015). Grain size also has a strong effect on the interpretation of geophysical observations such as seismic attenuation (Dannberg et al., 2017; Jackson et al., 2002; Tan et al., 1997) and electrical conductivity (Pommier et al., 2018; ten Grotenhuis et al., 2004). The grain size in the Earth's mantle is controlled by a few factors: grain-size reduction via dynamic or static recrystallization, grain growth and phase transitions (Solomatov and Reese, 2008). When considering deeper parts of the upper mantle (depths > 200km), static and dynamic recrystallization becomes less important as shear stresses decrease and diffusion creep arguably becomes the main deformation mechanism (Karato, 1992). Thus, grain growth is likely the most important factor controlling grain sizes in the deep upper mantle. However, in the absence of a grain-size reduction mechanism, grain growth might favour dislocation creep. The balance of these mechanisms at conditions of the Earth's deep upper mantle is not well understood and must be further investigated.

Grain growth is a mechanism driven by the need to minimize energy in aggregates through the migration of grain boundaries. The major forces driving grain boundary migration are interfacial and strain energy. Grain boundaries have high energy in comparison to crystal lattices. Thus, grains grow to minimize this excess energy by decreasing their surface-area-to-volume-ratio. This process is called normal grain growth (Atkinson, 1988; Burke and Turnbull, 1952; Evans et al., 2001; Humphreys and Hatherly, 2004), which leads to a uniform microstructure where the shape of the grain-size distributions is time-invariant (Atkinson, 1988). Normal grain growth is commonly described by the empirical equation:

$$d^p - d_0^p = kt \qquad (1)$$

where *d* and *d*$_0$ are the grain sizes at time = *t* and 0, respectively, n is the grain growth exponent and k is a rate constant given by:

$$k = k_0 \exp\left(-\frac{E^* + PV^*}{RT}\right) \qquad (2)$$

where $k_0$ is a material-dependant pre-exponential constant, $E^*$ the activation energy, $P$ the pressure, $V^*$ the activation volume, $R$ the gas constant and $T$ the temperature. Values for the grain growth exponent, *p*, are dependent on the mechanism of grain growth, usually varying between 1 and 4 for metals (Atkinson, 1988; Burke and Turnbull, 1952). For olivine aggregates, values for *p* were found to be between 2 for monomineralic olivine aggregates (e.g., Karato, 1989) and 4 for aggregates containing mixtures of olivine and pyroxene (e.g., Nakakoji and Hiraga, 2018). These results are consistent with those observed in metals, indicating grain-boundary migration in a pure system and a mixture of coalescence (Ostwald ripening) and grain-boundary migration in polymineralic aggregates as the main mechanisms controlling grain growth kinetics. Ostwald ripening is driven by differences in grain-boundary curvature: grains with a greater curvature (larger surface-area-to-volume-ratio, smaller grains) are consumed while grains with a lower curvature (lower surface-area-to-volume-ratio, larger grains) grow (Atkinson, 1988). In other words, the total grain boundary energy can be reduced when the total grain boundary area is reduced, which is, of course, also the driving force that leads to the simplified growth laws stating that grain boundaries migrate towards the centre of curvature. Notably, no relation between curvature and grain boundary migration velocity can be established (Bhattachary et al., 2021).

Particles of secondary phases might lead to modifications in grain growth of the major phase (matrix) by exerting a retarding force on the migrating boundaries (Humphreys and Hatherly, 2004). In other words, the interaction of a secondary phase particle with the grain boundary modifies the energy of this grain boundary. This process is known as Zener pinning, which depends on the properties of the moving boundary, such as its energy and mobility, as well as the properties of the second-phase particles, such as their size, shape and distribution (Nes et al., 1985). When pinning occurs, a balance exists between the forces leading to grain-boundary migration and those counteracting it due to pinning. This leads to a relation between the grain size of the primary phase ($d_I$) and of the second phase ($d_{II}$) that can be generalized as (Smith, 1948):

$$\frac{d_I}{d_{II}} = \frac{\beta}{f_{II}^z}, \qquad (2)$$

where $f_{II}$ is the volume fraction of the second phase, and $\beta$ and $z$ are theoretically or empirically calculated constants. For olivine and enstatite aggregates, Hiraga et al. (2010) found values of $\beta \approx 0.74$ and $z \approx 0.59$. Another less frequently observed phenomenon related to second-phase particles is the matrix's abnormal or discontinuous grain growth, i.e., the major phase (Hillert, 1965), which is restricted by the presence of second phases. Additionally, abnormal grain growth can occur due to anisotropic grain-boundary mobility or anisotropic grain-boundary energy (Rollett and Mullins, 1997).

Grain growth of olivine, the main phase in the Earths' upper mantle (Ringwood, 1970), has been investigated by several authors (Table 1). Karato (1989) investigated the grain growth of reconstituted San Carlos olivine aggregates at pressures of 0.1 MPa to 1 GPa, at dry and water-saturated conditions. He found that pores act as pinning particles at lower pressures, inhibiting grain growth. The same effect was found for water, that, when in excess (i.e., in free-water), fills the pores and also act as pinning particles. At lower concentrations, however, water was found to promote grain growth. Nichols and Mackwell (1991) investigated the grain growth of San Carlos olivine at atmospheric pressure, for varying oxygen fugacity. They found that the grain growth rate increased for increased oxygen fugacity. Faul and Scott (2006) studied the effect of melt in the grain growth kinetics of sol-gel olivine and they found that an increase in melt content led to a decrease in the grain growth rate. This suggests that melt inhibits grain growth of olivine in partially molten aggregates. Ohuchi and Nakamura (2007a, 2007b) analysed the grain growth of sol-gel forsterite in the forsterite-diopside and forsterite-diopside-water systems. They found that abnormal grain growth of forsterite was abundant when the secondary phase content (diopside) was less than 20 vol.%. Hiraga et al. (2010), Tasaka and Hiraga (2013) and Nakakoji and Hiraga (2018) investigated the grain growth of vacuum-sintered forsterite aggregates with enstatite fractions from 0 to 0.42. They found that the increase in the second-phase content reduced the rate of forsterite grain growth due to Zener pinning. Nakakoji and Hiraga (2018) further concluded that grain-boundary diffusion is a common mechanism responsible for grain growth and diffusion creep for olivine-dominated rocks. Speciale et al. (2020) investigated the grain growth of olivine from the Balsam-Gap dunite after static annealing, during deformation (dynamic recrystallization) and after deformation (static recrystallization). Surprisingly, they found that the grain growth rate of previously deformed olivine is significantly lower than that of undeformed olivine. Recently, Zhang and Karato (2021) investigated the grain growth of olivine aggregates, at pressures from 3-12 GPa. They found that grain growth kinetics of pure olivine aggregates is not strongly affected by pressure.

With the exception of Zhang and Karato (2021), all these above-summarized studies on normal olivine grain growth were conducted at relatively low pressures (up to 1.4 GPa) or with very low pyroxene fractions. However, olivine dominates the lithologies of the upper mantle to pressures of 14 GPa, with pyroxene as a common second phase (e.g. Ringwood, 1970). Thus, its influence should also be considered. Here we investigated grain growth kinetics of olivine in aggregates containing 6 and 13 vol. % of pyroxene (corresponding to dunite and harzburgite, respectively) at temperatures spanning from 1323 K to 1793 K and pressures from 1 GPa to 12 GPa. These parameters cover mineralogy, pressure and temperature conditions found in most of the Earth's upper mantle.

Table 1: Summary of experimental conditions for previously reported olivine grain growth experiments and the present study. P is pressure in GPa, T temperature in K, t experimental duration in hours and fPx is the pyroxene fraction.

| Starting material | P (GPa) | T (K) | t (hours) | $fO_2$ / Buffer | Water content | Melt content | Porosity (vol. %) | $f_{Px}$ | Study |
|---|---|---|---|---|---|---|---|---|---|
| San Carlos olivine | $10^{-4}$-1 | 1473-1673 | 0.5-200 | $10^{-5}$ Pa, Fe-Wüstite buffer | water free, water saturated | small amount (1) | 0 - 4.8 (2) | - | Karato (1989) |
| San Carlos olivine | $10^{-4}$-1 | 1473-1673 | 10-200 | $10^{-4}$ - $10^{-11}$ atm | water free | - | <5 | very small amount (3) | Nichols and Mackwell (1991) |
| Sol-gel olivine | 1 | 1523-1723 | 2-700 | C-CO buffer | water free | 2wt%, 4wt% | - | small amount | Faul and Scott (2006) |
| Sol-gel olivine | 1.2 | 1473 | 2-763 | Ni-NiO buffer | water free | - | <1 | 0.1 - 0.9 | Ohuchi and Nakamura (2007a) |
| Sol-gel olivine | 1.2 | 1473 | 1.5-763 | Ni-NiO buffer | water saturated | - | <1 | 0.1 - 0.9 | Ohuchi and Nakamura (2007b) |
| Vacuum sintered Forsterite | $5 \times 10^{-12}$ | 1633 | 0-50 | vacuum | water free | - | - | 0 - 0.42 | Hiraga et al. (2010) |
| Vacuum sintered Forsterite | $10^{-4}$ | 1533 – 1633 | 0-100 | - | water free | - | - | 0.01 - 0.97 | Tasaka and Hiraga (2013) |
| Vacuum sintered Forsterite | $10^{-4}$ | 1322 - 1669 | 500 | - | water free | - | - | 0.2 | Nakakoji and Hiraga (2018) |
| Balsam-Gap Dunite | 1.4 | 1373 - 1473 | 0.2-24.7 | Ni-NiO buffer | <50 ppm wt. $H_2O$ | - | <1 | <0.01 | Speciale et al. (2020) |
| San Carlos Olivine | 3-10 | 1273 - 1773 | 1-20 | Ni-NiO buffer | 9 - 54 ppm wt. $H_2O$ | - | - | 0.015 | Zhang and Karato (2021) |
| Sol-gel olivine | 1 - 12 | 1323-1793 | 12 - 72 | Ni-NiO buffer | <50 ppm wt. $H_2O$ | - | <0.5 | 0.06, 0.13 | This study |

[1]Glassy phases were found by the author, [2]Estimated from density measurements, [3]SiO2 originated from abrasion with agate mortar

# Methods

## Sample preparation

Here we studied grain growth in olivine and pyroxene aggregates fabricated through a solution-gelation method. The solution-gelation (sol-gel) method is an effective process to create chemically pure and homogeneous solids (Edgar, 1973; Hench and West, 1990). The procedure used here for olivine sol-gel synthesis is similar to the one previously described by Jackson et al. (2002). The precursors used as source of $SiO_2$, MgO and FeO were respectively tetraethyl orthosilicate (TEOS, $Si(OC_2H_5)_4$, *Sigma-Aldrich*, purity ≥ 99.0%), magnesium nitrate hexahydrate ($Mg(NO_3)_2 \cdot 6H_2O$, *Roth*, purity ≥ 99.999%) and Iron (III) nitrate nonahydrate ($Fe(NO_3)_3 \cdot 9H_2O$, *Sigma-Aldrich*, purity ≥ 98%).

Two different batches were fabricated with different amounts of TEOS, creating $SiO_2$ in excess to produce 6 vol.% (FSG4 batch) and 13 vol.% of pyroxenes (FSG5 batch). The reactants were dissolved in ethanol and gelation was reached by adding $NH_4OH$ (*Sigma Aldrich*, 25% $NH_3$). The gel was dried at increasing temperatures up to 773 K in air. The resulting powder was grounded and pelletized. The green body was then sintered in a gas mixing furnace with a controlled oxygen fugacity between the F-MQ and I-W buffers, while increasing temperature from 700 to 1673 K at a rate not higher than 300 K/hour. The sample was sintered at 1673 K for 8 hours and then slowly quenched by turning off the furnace power and waiting until the temperature reached 973 K, when the sample was removed from the furnace. The sintered olivine was reground, and the resulting powder pressed into a $Ni_{80}Fe_{20}$ capsule, filled at the top and bottom with a thin layer of NiO, to buffer the oxygen fugacity to the Ni-NiO buffer. The sample was kept in an oven at 423 K for at least 1 day before the capsule was weld shut. The sample was subsequently hot-pressed in a piston-cylinder apparatus at 0.7 GPa and 1473 K for 2 hours, yielding the starting material for all experiments. For the grain growth experiments in the piston cylinder, the starting material remained in the piston cylinder and the pressure and temperature were adjusted to the desired experimental conditions and held for the experimental duration. In preparation for the multianvil experiments, the sample pieces were cored out of the starting material and fired at 1273 K for 1 hour in a gas-mixing furnace. The firing of the samples at the respective temperature and time does not cause noticeable grain growth. These firing and annealing steps ensured that olivine aggregates were kept dry during the experiments, as water might affect olivine grain growth (Karato, 1989; Ohuchi and Nakamura, 2007b).

## Grain growth experiments

The grain growth experiments at 1 GPa were performed in a piston-cylinder apparatus and the experiments at higher pressures in a multianvil press. The assemblies were designed to reduce any deviatoric stress on the sample during compression. The piston-cylinder experiments were conducted using a 19 mm talc/Pyrex assembly (Figure 1a). The temperature was monitored during the experiments with an S-type (90% Pt/10% Rh–Pt) thermocouple. The experiments were performed by adjusting pressure to the target pressure, heating the sample to the target temperature at a 100 K/minute rate and maintaining these conditions for the experimental duration. The samples were quenched by reducing the current in the sample heater, to achieve 300 K temperature reduction per minute. This step reduces thermal shock in the sample and subsequent fracturing of grains. The pressure was reduced over 8 hours. The multianvil experiments were performed using second-stage WC anvils of 11 mm truncated edge length, acting on a $Cr_2O_3$-doped MgO octahedra (Figure 1b) with an edge length of 18 mm. The experiments in the multianvil apparatus were performed in analogy to the piston-cylinder experiments, except that longer decompression duration of at least 12 hours were necessary. The temperature in the multianvil experiments was monitored using a D-type (97% W/3% Re–75% W/25% Re) thermocouple. Uncertainties in pressure for the 19 mm piston-cylinder assembly are in the order of 0.02 GPa and thermal gradients are approximately 25 K within the sample at our experimental conditions (Watson et al., 2002). Uncertainties in pressure for the 18/11 multianvil assembly are around 0.5 GPa and thermal gradients in the order of 40 K within the sample at our experimental conditions (Walter et al., 1995). No correction was applied for the pressure effect on the electromotive force (emf) of the thermocouple, although for experiments at higher pressures (e.g. 7-12 GPa) temperatures of 20K to 40K higher than measured might have been reached (Nishihara et al., 2020).

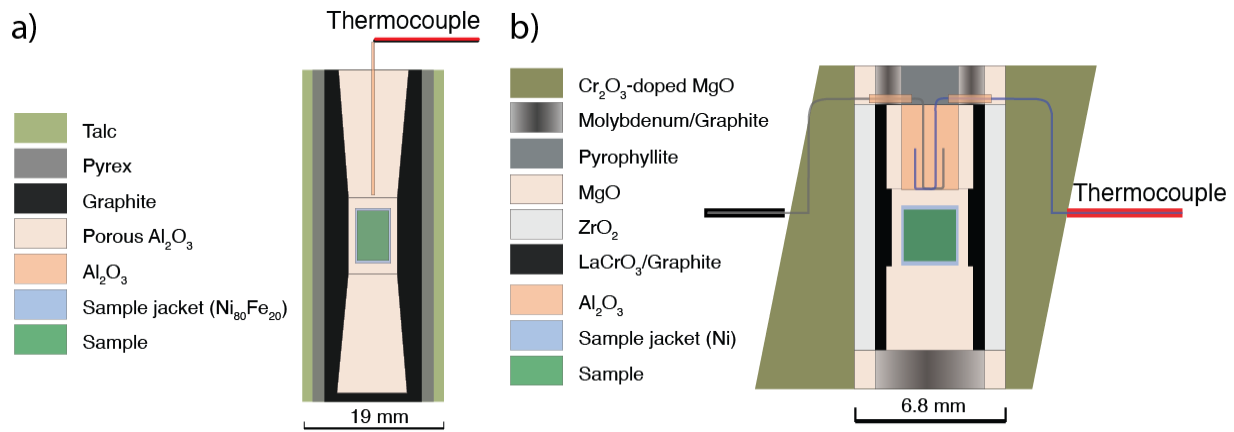

Figure 1: Cross-section diagrams of assemblies used in the a) piston cylinder and b) multianvil experiments.

Analytical techniques and grain-size measurements

Grain sizes were obtained from Electron Backscatter Diffraction (EBSD) data. EBSD data were collected with an *EDAX DigiView 5* EBSD detector mounted onto an *FEI Scios Dual-Beam* scanning electron microscope. For EBSD data acquisition we used an acceleration voltage of 20 to 30 keV and beam current of 3.2 to 6.4 nA. The EBSD data was acquired with the *EDAX TEAM™* software. The step size chosen for EBSD measurements for each sample is at least 10 times smaller than the olivine's average grain size and ranged from 0.1 to 0.5 μm (see Table 2). Thus, the uncertainty in grain size measurement of olivine for each sample is expected to be smaller than 10% of its average grain size. The uncertainty is proportionally larger for grains smaller than average. Pyroxene grains are on average 1.7 (13 vol.% Pyroxene) and 2.1 (6 vol.% Pyroxene) times smaller than olivine grains (see Table 2). Uncertainties in the EBSD measurements of pyroxene grains are higher as pyroxene grains have, on average, smaller grain sizes. Grains smaller than 10 times the step size in our EBSD measurements were excluded. Therefore, statistical analysis of pyroxene grain size is hindered in our study.

The EBSD data was cleaned using the *EDAX OIM Analysis™* software. First, electron backscatter patterns were re-indexed using the neighbour pattern averaging (NPAR) method (Wright et al., 2015). Subsequently, one iteration of grain dilation processing was applied, where pixels with low indexing quality, given by the confidence index, are assigned to the orientation of neighbours with a higher indexing quality. Olivine presents a pseudo-symmetry misindexing which correlates to a misorientation of 60° around the [100] axis (e.g., Marquardt et al., 2017). A pseudo-symmetry correction was therefore applied by merging neighbouring grains sharing a boundary with a misorientation axis of [100] and angle of 60º, keeping the orientation of the largest grain. Lastly, one more step of grain dilation was performed. Examples of the raw data and the effect of the cleaning steps are shown in Figure S.2.1. EBSD data was analysed with *MTEX* (Hielscher and Schaeben, 2008). Grains were defined as bounded regions where the misorientation angle exceeds 20°. This choice is based on the critical misorientation angle for the occurrence of dislocation arrays in olivine, as observed by Heinemann et al. (2005). Grains containing less than 20 indexed pixels were not considered.

Grain size was measured with three different methods: i) the largest dimension between any two vertices in a grain, which is the standard diameter function in MTEX, ii) by the equivalent diameter of a circle with the calculated grain's area, and iii) by the mean intercept length (MIL) using a rectangular grid. In the MIL method, the average grain size is given by the ratio between the length of a line and the number of grains intercepted by it. This method usually measures grain sizes in micrographs (e.g., Karato, 1989). We adapted this method for use with EBSD measurements of a single phase. Although the methods provide different results, the results are proportionally correlated (see Table 2), with the first method (largest dimension) providing grain sizes on average 1.31 ($\sigma = 0.02$) and 1.6 ($\sigma = 0.1$) times larger than the second (equivalent diameter) and third (mean intercept length) methods, respectively. Theoretical and experimental studies of normal grain growth in ceramics (e.g., Mendelson, 1969) describe a ratio of 1.6 between the average grain size and the mean intercept length, which indicates that the largest dimension method better describes the real grain size dimensions. Therefore, measurements of grain size, *d*, presented here are obtained using the first method, which also allows faster computation. We did not use any conversion factor. We highlight, however, that caution should be employed when comparing absolute grain size values between different studies (e.g., Hansen et al., 2011; Heilbronner and Bruhn, 1998). We also present the log-normal distributions fitted to grain size populations, which are often used to describe grain-size distributions (e.g., Faul and Scott, 2006; Mendelson, 1969; Tasaka and Hiraga, 2013).

Major and minor element chemistry was obtained from an electron microprobe equipped with a wavelength-dispersive spectrometer. The data was collected with an electron-beam voltage and current of 15 keV and 15 nA, respectively. Counting time was 20 seconds per element peak acquisition and 10 seconds for background collection. The water content was measured by unpolarized Fourier-transform infrared spectroscopy (FTIR), under atmospheric conditions. Doubly polished samples of 200 µm thickness were used. The spectra were obtained using an aperture of 100 µm and a resolution of 2 cm$^{-1}$. The spectrum baseline was fitted to a spline curve estimated using the *MATLAB*'s function *msbackadj*.

## Results

### Starting material: Microstructure, chemistry, and water content

The starting material for the grain growth experiments was characterized for its chemistry, microstructure, and water content. The chemical composition of the starting material (FSG4 and FSG5 batches) is exhibited in Table S.3.1. Olivine and pyroxene have a composition of Fo$_{90}$ and En$_{90}$, respectively. Pt, Al and Ni impurities were found at concentrations close to their expected detection limit (smaller than 0.05 wt.%). Pt and Al likely originate from the crucible and furnace used during the sintering and Ni from the capsule used in the annealing experiments. The microstructure of the starting material shows olivine and pyroxene grains uniformly distributed throughout the sample at a scale (Figure 2a). The grain size of olivine is generally larger compared to pyroxene. Furthermore, in a smaller scale, olivine grains surrounded by other olivine grains are usually larger than the olivine grains surrounded by pyroxene grains (Figure 2b and Figure 2c, respectively). Pyroxene grains regularly show nm-sized twinning (Figure 2d), causing the indexing of EBSD patterns of pyroxene to be difficult for the starting material. Note, that such twins, however, were not observed after the grain growth experiments at higher pressures (i.e., P ≥ 1 GPa). These polysynthetic twins are likely related to a transition between orthoenstatite (Pbca) and clinoenstratite (P2$_1$/c) (Ohashi, 1984), that occurs at low temperatures (T < 873 K) during quenching. The grain-size population shows a narrow log-normal distribution with average grain size (*d*) at 2.6 µm and 3.1 µm for batches FSG4 and FSG5, respectively (Figure 2e). At a larger scale (several 100 mm), the microstructure appears homogenous (Figure 2a), although, at a smaller scale, the grain size varies with the distribution of pinning particles (pyroxene grains, compare Figure 2b to 2c). The porosity of the samples was estimated from secondary electron imaging of the starting material. The pores (or inclusions) are mainly found at the grain boundaries and, rarely, in a grain's interior (Figure 2d). The pore area, calculated for different samples, averages to less than 0.5%. Because grains might be plucked out during grinding and polishing of the sample, leaving holes that appear as porosity, the actual porosity may be significantly less. No melt was observed in any of the samples. Cold compression of samples using solid pressure media might induce shear stresses within the sample (e.g., Liebermann and Wang, 2013; Rubie et al., 1993). The accumulated stresses in the sample might lead to modifications in the grain sizes. The effect of grain size modification because of cold compression was evaluated through an experiment in the multianvil apparatus at 10 GPa without heating. The results of this experiment, Z1993, are compared to its starting material, sample FSG4, in Figure 2e. The similar grain-size distribution between FSG4 and the Z1993 samples, indicates that no grain-size reduction during cold compression and decompression occurred. Similarly, neither the starting material nor experimental run products from the piston-cylinder based experiments show signs of intergranular fracturing or grain size reduction (Figure 2a-d). The water content was measured for the starting material after hot-pressing in the piston cylinder and after annealing experiments at 1 GPa, 7 GPa and 10 Gpa. The measured water content for all samples falls below the detection limit of approximately 50 ppm (Figure S.2.2), indicating that the aggregates were dry (see Figure 5 of Faul & Jackson (2007) for an FTIR spectrum comparison of dry and wet olivine aggregates).

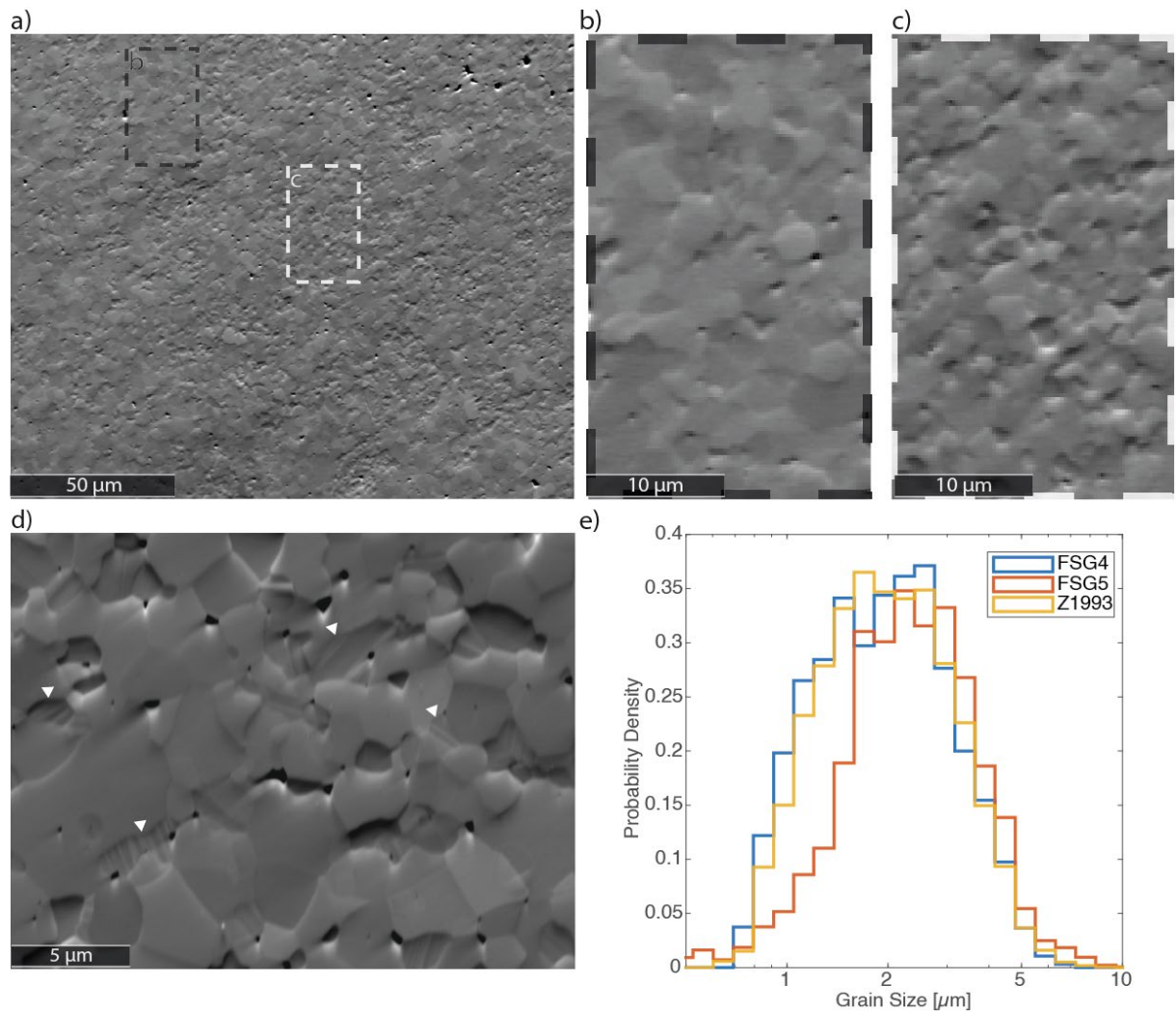

Figure 2: Starting material microstructure and grain-size distribution: Forescattered electron images of sample FSG-4 (a-d) show olivine grains elevated with respect to smaller often twinned pyroxene grains. Holes appear dark with a bright shadow. a) Olivine grains not in contact with pyroxene (black rectangle, magnified in b)) are on average larger than those surrounded by pyroxene (white rectangle, magnified in c)). d) High magnification image from the same sample shows that pores/holes are mostly distributed at the grain boundaries. Pyroxene often presents lamellar twinning, as indicated by the white arrowheads. e) Grain-size distribution of starting material of batches FSG4 (6 vol.% Px) and FSG5 (13 vol.% Px) and of sample Z1993 (High-pressure experiment (10 GPa) without heating).

Grain-size evolution

Experimental conditions were chosen to allow investigation of the effect of time, temperature, and pressure on grain growth of olivine. Figure 3 shows the microstructure after the experiments and Figure 4 shows the grain-size distributions. Histograms of the grain-size distributions are shown in Figure S.2.2. In the series of experiments for different annealing times (Figures 3a and 4a-c) we analysed the effect of annealing time on grain growth with experiments performed at 1673 K, pressures of 1 and 7 GPa and experimental durations of 8 to 72 hours. In the temperature series (Figures 3b and 4d-f) we evaluate the effect of temperature on grain growth with experiments performed at an experimental duration of 24 hours, pressures of 1 and 7 GPa and temperatures of 1323 K to 1793 K. In the pressure series (Figures 3c and 4g-i) we analysed the effect of pressure on grain growth with experiments performed at an experimental duration of 24 hours, temperature of 1673 K and pressures between 1 and 12 GPa for 6 and 13% Px vol.%. Table 2 shows a summary of the obtained experimental data.

Table 2: Experimental Data: P is pressure in GPa, T temperature in K, t experimental duration in hours, d average grain size in µm, dEQ the equivalent diameter of olivine grains in µm and dMIL the mean intercept length in µm. MoF$_{IT}$, µ$_{FIT}$ and σF$_{IT}$ are the mode, mean and standard deviation of the lognormal fit to the grain-size distribution, respectively. fPx is the pyroxene fraction as measured by EBSD and dEQ Px is the equivalent diameter of pyroxene grains in µm. n is the number of grains analysed for each sample and s is the maximum EBSD step size.

| Sample | Starting Material | P (GPa) | T (K) | t (h) | d (µm) | d$_{EQ}$ (µm) | d$_{MIL}$ (µm) | Mo$_{FIT}$ | µ$_{FIT}$ | σ$_{FIT}$ | fPx (EBSD) | d$_{EQ}$ Px (µm) | n | s (µm) |
|---|---|---|---|---|---|---|---|---|---|---|---|---|---|---|
| FSG4 | Sol-gel Ol (Fo$_{90}$) + 6%Px | 0.7 | 1473 | 2 | 2.57 | 1.91 | 1.44 | 1.91 | 0.85 | 0.45 | 0.01 | 1.11 | 1046 | 0.1 |
| FSG5 | Sol-gel Ol (Fo$_{90}$) + 13%Px | 0.7 | 1473 | 2 | 3.06 | 2.34 | 1.91 | 2.29 | 1.02 | 0.44 | 0.07 | 1.46 | 1578 | 0.2 |
| Z1993 | FSG4 | 10 | 298 | 1 | 2.62 | 1.94 | 1.47 | 1.96 | 0.87 | 0.44 | 0.02 | 1.29 | 3850 | 0.15 |
| Z1962 | FSG4 | 5 | 1673 | 24 | 15.39 | 11.9 | 10.05 | 9.76 | 2.59 | 0.56 | 0.06 | 4.37 | 4185 | 0.5 |
| Z1965 | FSG4 | 7 | 1673 | 24 | 9.09 | 6.91 | 5.75 | 6.44 | 2.09 | 0.47 | 0.02 | 3.78 | 22315 | 0.5 |
| Z1968 | FSG4 | 12 | 1673 | 24 | 8.48 | 6.45 | 5.52 | 6.07 | 2.02 | 0.47 | 0.01 | 3.52 | 21069 | 0.5 |
| A1178 | FSG5 | 1 | 1673 | 12 | 6.94 | 4.99 | 4.84 | 4.62 | 1.8 | 0.52 | 0.08 | 2.68 | 1550 | 0.5 |
| A1179 | FSG5 | 1 | 1673 | 72 | 6.1 | 3.09 | 4.43 | 3.92 | 1.66 | 0.54 | 0.1 | 1.84 | 2296 | 0.4 |
| A1182 | FSG5 | 1 | 1673 | 8 | 6.13 | 2.65 | 4.41 | 3.88 | 1.66 | 0.55 | 0.1 | 1.72 | 3335 | 0.25 |
| B1272 | FSG5 | 1 | 1673 | 24 | 6.49 | 2.42 | 4.61 | 4.33 | 1.73 | 0.51 | 0.09 | 1.64 | 4697 | 0.4 |
| B1273 | FSG5 | 1 | 1473 | 72 | 4.01 | 2.81 | 2.75 | 2.8 | 1.26 | 0.48 | 0.1 | 1.79 | 6087 | 0.3 |
| B1274 | FSG5 | 1 | 1473 | 24 | 3.45 | 2.68 | 2.25 | 2.55 | 1.13 | 0.44 | 0.08 | 1.79 | 7146 | 0.2 |
| B1275 | FSG5 | 1 | 1323 | 24 | 3.16 | 2.57 | 2.01 | 2.36 | 1.05 | 0.44 | 0.11 | 1.74 | 8692 | 0.2 |
| B1276 | FSG5 | 1 | 1473 | 12 | 3.71 | 2.17 | 2.35 | 2.67 | 1.19 | 0.46 | 0.09 | 1.56 | 9717 | 0.25 |
| Z2032 | FSG5 | 10 | 1673 | 24 | 3.46 | 2.54 | 2.16 | 2.69 | 1.16 | 0.41 | 0.09 | 1.69 | 14505 | 0.2 |
| Z2033 | FSG5 | 7 | 1473 | 24 | 3.34 | 3.16 | 2.1 | 2.6 | 1.12 | 0.41 | 0.08 | 2.16 | 8450 | 0.2 |
| Z2034 | FSG5 | 7 | 1473 | 12 | 2.85 | 8.84 | 1.67 | 2.31 | 0.98 | 0.37 | 0.07 | 3.89 | 11578 | 0.2 |
| Z2035 | FSG5 | 7 | 1323 | 24 | 3.3 | 4.22 | 2.09 | 2.53 | 1.1 | 0.42 | 0.09 | 2.58 | 9587 | 0.2 |
| Z2047 | FSG5 | 7 | 1673 | 12 | 4.1 | 2.79 | 2.57 | 3.11 | 1.32 | 0.43 | 0.12 | 1.79 | 3808 | 0.2 |
| Z2049 | FSG5 | 7 | 1793 | 24 | 11.42 | 5.29 | 7.76 | 7.22 | 2.28 | 0.55 | 0.13 | 2.87 | 5112 | 0.5 |
| Z2051 | FSG5 | 7 | 1673 | 72 | 5.44 | 4.72 | 3.89 | 3.45 | 1.54 | 0.55 | 0.12 | 2.54 | 5862 | 0.25 |
| Z2060 | FSG5 | 7 | 1673 | 24 | 3.62 | 3.72 | 2.25 | 2.8 | 1.2 | 0.41 | 0.08 | 2.2 | 4484 | 0.2 |

| Z2062 | FSG5 | 7 | 1673 | 48 | 4.83 | 4.74 | 3.31 | 3.37 | 1.45 | 0.48 | 0.08 | 2.33 | 5039 | 0.3 |

The grain-size distribution resulting from the series of experiments at different annealing times at 1 GPa (Figure 4a) demonstrates a flattening and broadening compared with the grain-size distribution of the starting material (Figure 2e). The average grain size for the 8-hour experiment is identical to the average grain size of the 72-hour experiment ($d$ = 6.1 µm). Similar flattening and broadening of the grain-size distribution occur only after 72 hours for experiments conducted at 7 GPa (Figure 4b). The temperature series of experiments at 1GPa (Figure 4d), demonstrates grain-size distributions similar to the starting material ($d$ = 3.1 µm) for the samples annealed at 1323 K ($d$ = 3.2 µm) and 1473 K ($d$ = 3.7 µm). For the experiment at 1673 K, the grain-size distribution is flattened, with its peak shifted towards larger grain sizes ($d$ = 6.5 µm). At 7 GPa (Figure 4e), a comparable effect is only observed for the experiment done at 1793 K ($d$ = 11.4 µm).

The pressure series of experiments for samples with 6 vol.% of pyroxene (Figure 4g), demonstrates similar distributions for experiments performed at 7 and 12 GPa, while, in comparison, the experiment performed at 5 GPa shows more spread in the grain-size distribution, which is also shifted towards larger grain sizes. The experiments performed with aggregates containing 13 vol.% of pyroxene (Figure 4h) present a similar effect. While the experiments done at 7 and 10 GPa exhibit similar grain-size distributions, larger grain sizes resulted at experiments performed at 1 GPa.

Figure 5b displays the minimum distance between pyroxene grains (see scheme in Figure 5a) normalized by the mean grain size. In our experiments, we observe that this distance increases for increasing pressure. The mean normalized distance between pyroxene grains at 1 GPa is approximately 20% smaller than at 7 GPa and 12 GPa (0.99, 1.23 and 1.25 respectively). Figure 5c demonstrates the local grain-size distribution of olivine grains (FSG5 sample) as a function of the ratio between the number of pyroxene neighbours and all neighbours. Olivine grains with more pyroxene neighbours are generally smaller than olivine grains mainly in contact with other olivine grains. This effect is independent of pressure, temperature, and duration of experiments (see Figure 2 and 3).

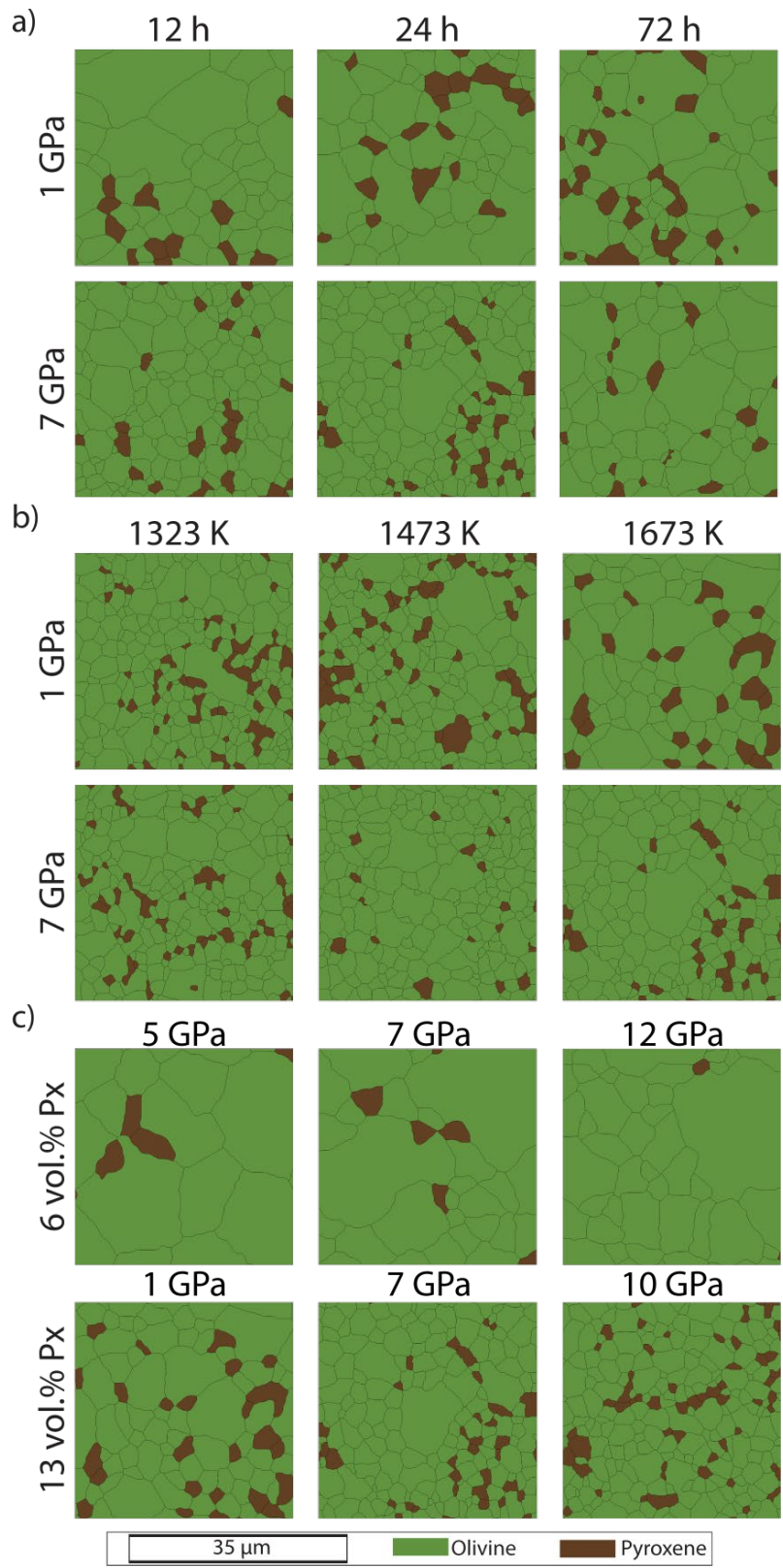

Figure 3: Microstructure evolution after grain growth experiments. Phase maps show a small representative subsection of the areas analysed: a) Time series of experiments: All experiments in this series were performed at 1673 K for aggregates of olivine plus 13 vol.% pyroxene. b) Temperature series: All experiments in this series were performed for 24h for aggregates of olivine plus 13 vol.% pyroxene. c) Pressure series: All experiments were performed at 1673 K and 24h. Olivine grains are coloured green and pyroxene brown. Plot areas are 35 x 35 μm.

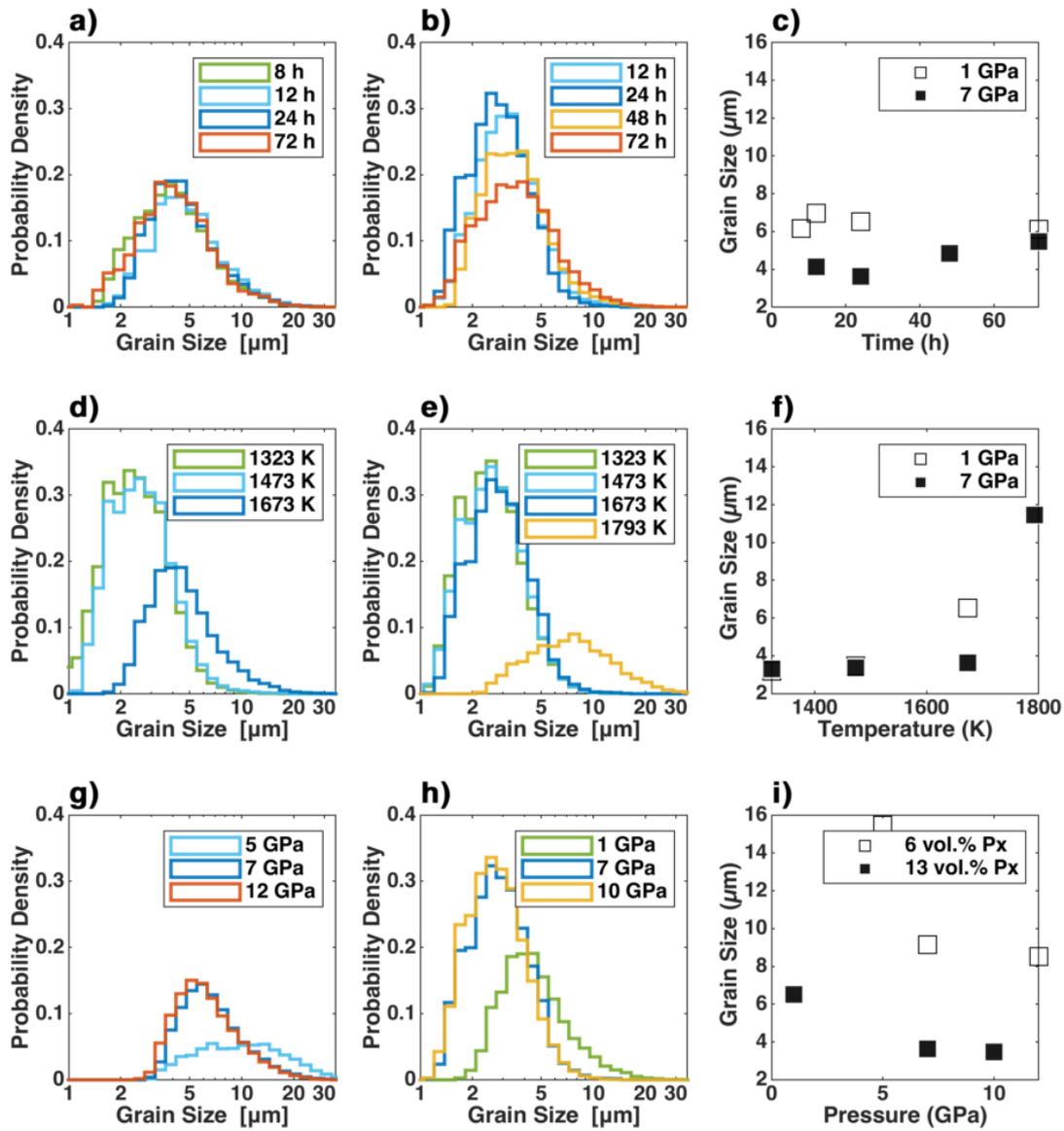

Figure 4: Grain-size distributions: The upper row (a-c) shows a time series of experiments performed at 1673 K and pressures of a) 1 GPa and b) 7GPa. c) Average grain size, as a function of time. The middle row (d-f) shows the temperature series of experiments performed for 24 hours at pressures of d) 1 GPa and e) 7GPa. f) Average grain size as a function of temperature. The bottom row (g-i) shows the pressure series of experiments performed at 1673 K for 24 hours for samples containing g) 6 vol.% and h) 13 vol.% of pyroxene. i) Average grain size as a function of pressure for samples with a Pyroxene content of 6 vol.% and 13 vol.%.

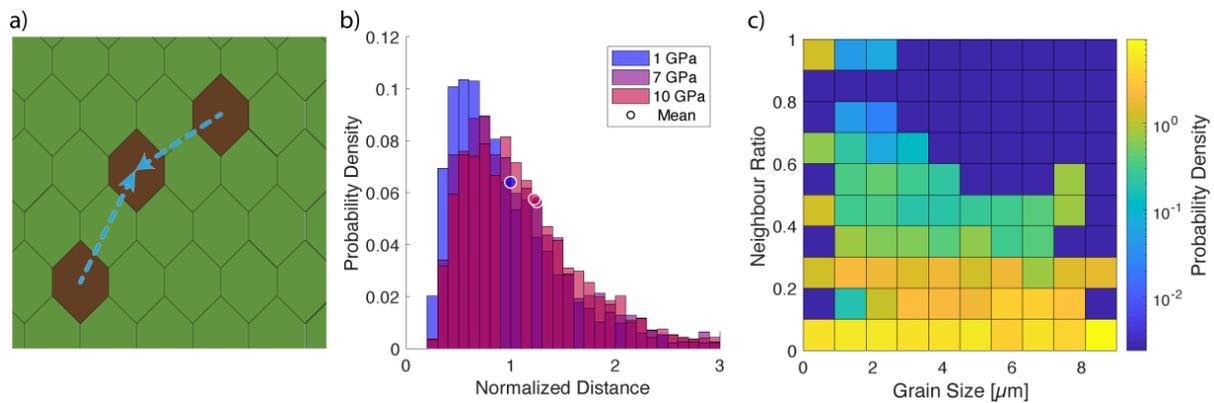

Figure 5: Effect of pyroxene on local grain-size distribution: a) Schematic of an aggregate containing olivine (coloured green) and pyroxene (coloured brown). Arrows indicate the distance between the centroid of pyroxene grains to the next nearest pyroxene grain. b) Minimum distance between pyroxene grains (as shown by the arrows in figure a) normalized by the average grain size. Experiments performed at 1673 K, 24h, and pressures of 1, 7 and 10 GPa. c) Olivine grain size as a function of the ratio between the number of pyroxene grains and all neighbour grains for sample FSG5 (starting material, 13 vol% Px). Each column is a probability density function.

## Discussion

Our results show that pressure has a significant effect on olivine grain growth. Figures 3a-b and Figures 4a-c demonstrate that grain growth is faster at 1 GPa compared to 7 GPa. For instance, at a temperature of 1673 K and annealing time of 24 hours, and 1 GPa grains grow rapidly ($d$ = 6.5 μm), while at the same temperature, annealing time and $P$ = 7 GPa, the grain-size distribution ($d$ = 3.6 μm) remains similar to its starting material ($d$ = 3.1 μm). At low temperatures (T≤ 1473 K) olivine grain growth is slow, irrespective of the applied pressure. Figures 3b and 4d-f show that the grain-size distribution of experiments performed at T≤ 1473 K and P = 7GPa (Z2033, Z2034, Z2035) is, within uncertainties, similar to their starting material.

The experiments at 1 GPa and 1673 K show almost no grain growth. We have no reason to exclude this experiment, so we included it in our fitting. We note that the minimal grain growth of the experiment at 1 GPa and 1673 K is inconsistent with experimental results at other pressures (7 and 10 GPa). This indicates that the normal grain growth law may not capture all processes that occur at grain boundaries. Notably, recent works on ceramics as well as silicates, including olivine, have proven that grain boundaries can change their structure (complexion or phase) as a function of temperature, pressure and chemistry (Furstoss et. al 2022, Cantwell et al. 2014, Kelly et al.2018, van Driel et al. 2020, Yokoi and Yoshiya 2018).

It is important to highlight that at pressures smaller than at least 1 GPa, pressure has an opposite effect than the one discussed here for aggregates containing pores. That is because the pressure in this range acts to reduce porosity; the olivine grain growth rate increases due to fewer pores acting as pinning particles (e.g., Karato, 1989). To test this observation, we performed grain growth experiments with vacuum-sintered forsterite (starting average grain size $d$ = 4.0 μm) +10 vol.% enstatite (Koizumi et al., 2010) at 1673 K during 24 h, at atmospheric pressure (0.1 MPa), using a tube furnace (HV806-HT) and using a piston-cylinder apparatus at 1 GPa (HV806-HP). The results reported in Table S.3.1 and Figure S.2.4 demonstrate that, after 24 hours, grains grow faster at 1 GPa ($d$ = 6.6 µm) compared to 0.1 MPa ($d$ = 5.0 µm). This demonstrates that even a small proportion of pores inhibits grain growth. Therefore, studies on grain growth and diffusion creep obtained at environmental or low pressures, even for a starting material with very low porosity, must be interpreted considering the pinning effect of pores.

Figures 3c and 4g-i demonstrate that the pressure effect on the grain growth rate is more pronounced for pressures between 1 and 7 GPa. For experiments containing 6 vol.% of pyroxene (Figure 4g), grains grow faster at 5 GPa ($d$ = 15.4 µm) than at 7 GPa ($d$ = 9.1 µm) and 12 GPa ($d$ = 8.48 µm). For experiments containing 13 vol.% of pyroxene (Figure 4h), grains grow approximately faster at 1 GPa ($d$ = 6.5 µm) than at 7 GPa ($d$ = 3.6 µm) and 10 GPa ($d$ = 3.5 µm).

The results presented in Figure 4i indicate that the effect of pressure on slowing down the grain growth of olivine increases with increasing pyroxene content. One possible explanation is that the grain growth rate of pyroxene is also reduced with increasing pressure. With a lower grain growth rate (for example, at pressures higher than 7 GPa), the pyroxene grains are kept far apart from one another (Figure 4b). This results in a longer diffusion path, for example of Si, through olivine grain boundaries, possibly reducing the coalescence of pyroxene grains (Nakakoji and Hiraga, 2018). Thus, at higher pressure, more pyroxene grains remain small. In fact, our results show that, at the same temperature, annealing duration and pyroxene content (1673 K, 24 h and 13 vol.% Px, respectively), the ratio between the number of pyroxene grains and their total area at 7 and 10 GPa (0.34 µm$^{-2}$) is more than twice that of at 1 GPa (0.16 µm$^{-2}$).

The larger the number of neighbouring pyroxene grains, the smaller is the olivine grain sizes (Figure 5c, see also Figures 2 and 3). Furthermore, grain growth is expected to be affected by the grains' neighbouring phase via Zener pinning. (Smith, C.S. 1948, Hiraga et al. 2010, Tasaka and Hiraga, 2013, Solomatov et al. 2002). In a first approximation, Zener pining describes the effect of a secondary phase on grain growth. If Zener pining controls grain growth, the relation between the primary phase grain size and the secondary phase grain size times the volume fraction of the secondary phase remains constant during grain growth (Smith, C.S. 1948, Evans et al., 2001). As the solubility of the secondary phase in the primary phase doesn't change during the experiment, it is a constant. Similarly, the normalized distance between pyroxene grains is interesting because if the pyroxene grain size is proportional to the grain spacing, it indicates that Zener pinning controls grain growth. In contrast, we observe that the above ratio increases with increasing pressure from ~0.9, over ~1 to ~1.25. This indicates that the pyroxene grains are further apart at higher pressures, and consequently, Zener pining is less effective at higher pressures.

It is not surprising that Zener pining, as described here, is only a first approximation and was developed for secondary phase particles orders of magnitude smaller than the primary grain size. Considering the samples with ~15% of 2$^{nd}$ phase, even with a relatively homogeneous distribution, not all matrix grains are pinned. (e.g. Figures 2 and 4). The matrix grains that are not pinned grow more than the pinned ones, which agrees with previous observations (e.g. Solomatov et al 2002, Ohuchi and Nakamura 2007, Karato 2008 p 238). Solomatov et al. (2002) demonstrated that this would lead to an asymptotic grain size evolution with time, similar to that observed in our experiments.

For example, experiments performed at 1 GPa, 1673 K and annealed for 8h (A1182), 12h (A1178), 24h (B1272) and 72h (A1179) show, within uncertainties, similar grain size distribution ($d$ = 6.5 ± 0.5 µm). This contrasts with a continuous increase in grain size with time in experiments with low (<1%) amounts of the second phase (e.g., Karato, 1989; Zhang and Karato, 2021)We conclude that the secondary phase, enstatite, affects grain growth but not to the full extent expected by Zener pinning. Notably, this will be of greater importance in multi-phase systems of the Earth's upper mantle.

Figures 4g-i demonstrate that the grain growth rate decreases considerably in a narrow range of pressure, between 5 and 7 GPa. Lithostatic pressures in this range correspond to depths of approximately 200 km. At this depth, a marked seismic discontinuity is observed (Lehmann, 1961, 1959), which has been argued to be related to a change in olivine main deformation mechanism, from dislocation creep to diffusion creep for increasing depths (Karato, 1992; Karato and Wu, 1993). Our data suggest that smaller grain sizes, resulting from the pressure effect on the grain growth, could be maintained at larger depths. This change in grain size might influence the variation in the dominant deformation mechanism in this region, facilitating grain-size sensitive mechanisms, such as diffusion creep, as depth and pressure increases.

Here we presented experimental evidence that the grain growth rate of olivine in aggregates containing olivine and pyroxene decreases for increasing pressures. Therefore, pressure affects the grain-boundary migration process in these aggregates. Grain-boundary diffusion was proposed to be the main process controlling grain growth and diffusion creep of olivine (Nakakoji and Hiraga, 2018). Grain growth in multi-phase aggregates is rate-limited by the removal, diffusivity and incorporation of the slowest moving ion (Farver and Yund, 2000; Karato, 2008, p. 238). Si is the slowest diffusing species in both the lattice and at the grain boundaries of olivine at T > 1473K (Farver and Yund, 2000). The Si grain-boundary diffusion coefficient of olivine was shown to decrease with increasing pressure (Fei et al., 2016). Consequently, we propose that the decrease in the grain growth rate of olivine can be explained by the decrease in the rates of Si grain-boundary diffusion with increasing pressure. This is further supported by the finding that:

1. An increase in pressure causes an increase in the melting temperature of olivine. In other words, at the same nominal temperature, the homologous temperature is reduced at high pressures. For example, the homologous temperatures for forsterite at 1623 K and pressures of 1 GPa, 7 GPa and 12 GPa are 0.72, 0.63 and 0.58, respectively (Davis and England, 1964; Ohtani and Kumazawa, 1981). Lower homologous temperatures would thus explain the reduced grain-boundary diffusion kinetics (e.g., Atkinson, 1985).
2. Grain-boundary energy anisotropy and the presence of specific grain boundaries with low mobility or with high solute segregation modify grain growth rates in metals and ceramics (e.g., Bäurer et al., 2013; Gottstein and Shvindlerman, 2009; Rheinheimer et al., 2015). Grain boundary plante populations are expected to be self-similar during grain growth in olivine but may change with pressure (e.g., van Driel et al., 2020; Yokoi and Yoshiya, 2018) and could potentially contribute to the decrease in the olivine grain growth rate reported here. While evaluating the grain boundary plane populations is beyond the scope of this contribution, we show in Figure S.2.5 that the misorientation angle and axis distribution changes for samples annealed at different pressures. Together with the fact that the population of olivine grain boundaries are also affected by deformation (Ferreira et al., 2021; Marquardt et al., 2015; Marquardt and Faul, 2018), this might indicate that different populations of grain boundaries are preferentially formed at different depths, which must be further investigated.

Commonly, the fitting of grain growth data is done sequentially by first fitting the grain growth exponent, then $k_0$, followed by activation enthalpy, energy, and volume (e.g. Speciale et al. 2020, Tsujino and Nishihara, 2009). However, the uncertainties for each fitting step are not propagated to the following steps. Although this procedure returns lower uncertainties for each parameter, it does not reflect the overall uncertainty. In other words, a range of values (e.g. k0, n and V*) allows for minimizing errors during the fit. To avoid reporting misleadingly low errors, we performed a global non-linear least-absolute residual fitting of our data at 1673 K and 13 vol.% Px to equations 1 and 2 (Figure 6). Data at different temperatures and pyroxene content were not included as it greatly increased fitting complexity. The activation energy, $E^*$, was fixed in the range [200, 700] kJ/mol, which is commonly observed for the grain growth of olivine (e.g., Faul and Scott, 2006; Nakakoji and Hiraga, 2018). The resulting fitting gives $k_0$ = 6.2 x $10^{-8}$ ($m^{4.18}s^{-1}$), *p* = 4.18, $V^*$ = 4.75 x $10^{-6}$ ($m^3$/mol) and $E^*$ = 617 (kJ/mol).

Comparison of our grain growth equation with previously published olivine grain growth data (Figure 6c) shows that our results at 1673 K, 7 GPa and 13 vol.% Px, and those from Tasaka and Hiraga (2013) at *P* = 0.1 MPa 9 vol.% En and extrapolated to 1673 K indicate smaller grain sizes than those found by Karato (1989, 2008, p. 241), Speciale et al. (2020) and Zhang and Karato (2021) extrapolated to 7 GPa. This difference is likely related to the low amount of pyroxene in the experiments of Karato (1989, 2008, p. 241) and Zhang and Karato (2021) when compared to our results and those of Tasaka and Hiraga (2013). The activation volume found in our study is effectively the same as found by Fei et al. (2016) for Si grain-boundary diffusion at high pressures (4.0 ± 0.7 x $10^{-6}$ $m^3$/mol). Although the activation energy we found (607 kJ/mol) is similar to the one found by Tasaka and Hiraga (2013) (640 kJ/mol), it differs significantly from the one found by Fei et al. (2016) (220 ± 30kJ/mol). The reason for this discrepancy is unknown, as the starting material of the experiments of Tasaka and Hiraga (2013) and Fei et al. (2016) is similar (vacuum-sintered Forsterite), apart from the Pyroxene content, which was demonstrated not to considerably affect the activation energy (Tasaka and Hiraga, 2013). The activation energy and activation volume found in our study are also similar to the ones found by Speciale et al. (2020) (620 kJ $mol^{-1}$ and 5 x $10^{-6}$ $m^3$/mol, respectively), even though a different grain growth mechanism might be operative in their experiments (i.e. strain-driven grain-boundary migration), at least in the first few hours after deformation was stopped. One of the main parameters influencing the extrapolation of grain growth equations to longer durations (e.g. 1 to 100 Ma) is the grain growth exponent *p*, which depends on the rate-limiting physical process controlling grain growth, e.g., whether pinning occurs or not. The grain growth exponent we obtained (*p* = 4.18) is similar to the ones obtained by Tasaka and Hiraga (2013) and Ohuchi and Nakamura (2007a) in experiments with similar second-phase content and higher than the experiments with lower second-phase content (*p* = [2, 3.2], Karato, 1989; Speciale et al., 2020; Zhang and Karato, 2021). At the conditions of our experiments, grain growth of the primary phase (e.g. olivine) is likely controlled by grain-boundary diffusion and coarsening is limited by Zener pinning (e.g., Evans et al., 2001; Hillert, 1965). Based on these observations, we understand that the grain sizes predicted using our best-fit parameters more closely approximate the evolution of grain sizes of olivine in the middle to deep upper mantle.

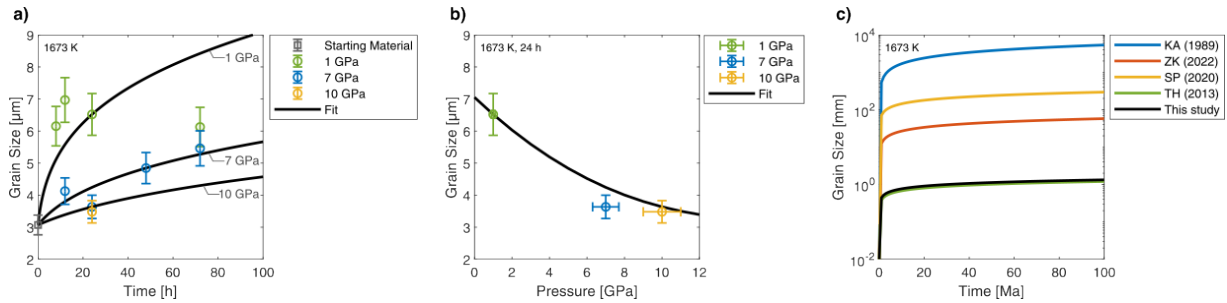

Figure 6: Fitting of grain growth data: a) Experimentally determined grain sizes (13 vol.% Px, *T*=1673 K, P=1, 7 and 10 GPa) and predicted grain sizes (curves) as a function of time. b) Experimentally determined grain sizes (13 vol.% Px, *t* = 24 h, *T*=1673 K) and predicted grain sizes (curve) as a function of pressure. Error bars show uncertainties in grain size and pressure determination. c) Extrapolation of predicted grain sizes to up to 100 Ma and comparison with existing studies: KA: (Karato, 1989), ZK: (Zhang and Karato, 2021), SP: (Speciale et al., 2020), TH: (Tasaka and Hiraga, 2013).

Our fitting results, however, carry uncertainty. The 95% confidence intervals for $k_0$, *p*, *V\**, and *E\** are respectively: [-3.84, 3.84] m$^{4.2}$s$^{-1}$, [-4.43, 12.78], [-2.32 x 10$^{-6}$, 1.18 x 10$^{-5}$] m$^3$/mol and [-8.61 x 10$^{11}$, 8.61 x 10$^{11}$] kJ/mol. The main reason for the large uncertainties in our fit is that a trade-off exists among the variables to be fitted. In order to demonstrate this trade-off, we calculated the root-mean-square error ($RMSE$) between predicted grain sizes ($\hat{d}$) and experimentally determined grain sizes ($d$), according to:

$$RMSE = \sqrt{\frac{\sum_{i=1}^{n}(\hat{d}_i - d_i)^2}{n}} \qquad (3)$$

The $RMSE$ is thus minimized when the difference between the predicted grain sizes and the experimentally determined grain sizes is minimized. The predicted grain sizes were obtained by fixing the activation energy, $E^*$, nd the grain growth exponent, *p*, within the range of previously reported values for grain growth in olivine aggregates: [200,600] kJ/mol and [2,5], respectively, and assuming values for $k_0$ and *V\**, to be within the range [$10^{-26}$, $10^{-4}$] m$^p$s$^{-1}$ and [1,8] x $10^{-6}$ m$^3$/mol, respectively. Figure 7 shows the resulting $RMSE$ as a function of $k_0$, and *V\** (Figures 7 a-c), the comparison between the experimental and predicted grain sizes (Figures 7 d-f) and the extrapolation of predicted grain sizes to geological times (Figure 7 g-i) for fixed $E^*$ = 200 kJ/mol and *p* = 2 (left column), 400 kJ/mol and *p* = 4 (middle column) and 600 kJ/mol and *p* = 5 (right column). Combinations among this range of $E^*$ d *p* are presented in Figures S.2.6 - S.2.8.  Figures 7 a-c demonstrates that within the investigated range of $E^*$ and *p*, similar $RMSE$ are obtained. Figures 7 d-f demonstrates that the predicted grain sizes fit similarly well to the experimentally determined grain sizes within the uncertainties of measurement. These results demonstrate that a trade-off exists among these parameters that would yield similar $RMSE$, and, therefore, would equally create predictions that closely fit the observations. Due to the large uncertainties in the determination of the grain growth parameters, especially of *p*, a large uncertainty exists when extrapolating the experimental data to a geological time scale. For example, Figure 7g-i demonstrates that, after 100 Ma, approximately three orders of magnitude difference in the final grain size is predicted considering *p* = 2 (Figure 7g) or *p* = 5 (Figure 7i).  We highlight that this is not exclusive to our dataset and the same can be demonstrated to other minerals, e.g. garnet (Ezad, 2019). As pointed out by Ezad (2019), even experiments performed over a month, which, at present, are extremely challenging to perform at high pressure and temperature, would not be sufficient to significantly reduce these uncertainties when extrapolating to geological-relevant residence times. Novel computational and experimental approaches are necessary to increase confidence when extrapolating between time scales.

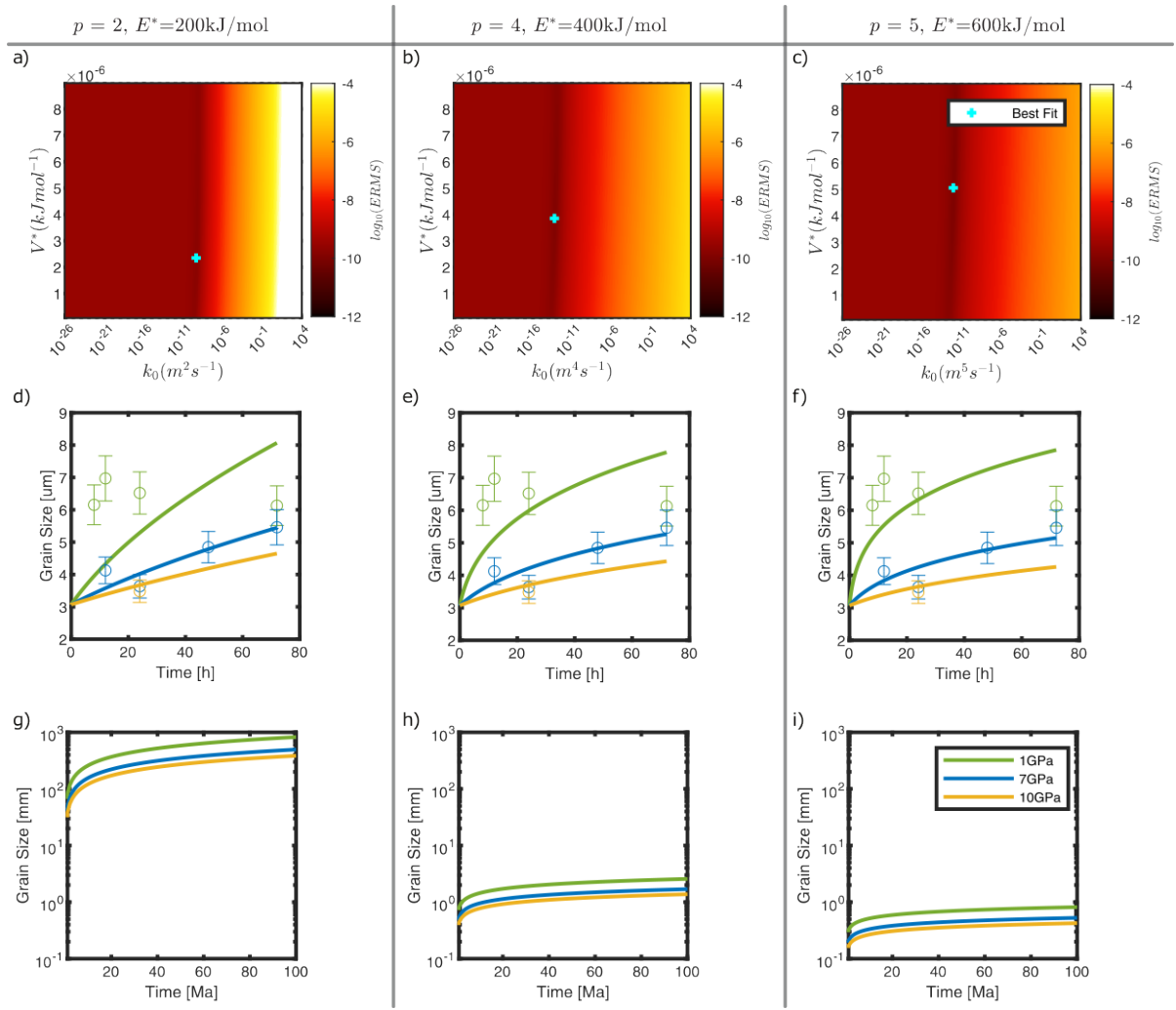

Figure 7: Uncertainties during fitting of grain growth data: a-c) Root-mean-squared error (*RMSE*) as a function of $k_0$ and $V^*$. Blue cross indicates the minimum *RMSE* for each figure (best fit). d-f) Comparison between experimentally determined grain sizes (13 vol.% Px) and predicted grain sizes (curves) considering the best-fit parameters above, and as a function of time, at P=1, 7 and 10 GPa. Error bars show uncertainties in grain size determination. g-i) Extrapolation of predicted grain sizes to up to 100 Ma. All plots are obtained considering temperature *T*=1673 K.

When extrapolating our data to conditions of the Earth, two other main uncertainties are the initial grain size, $d_0$, and the annealing time, $t$. Grain sizes of approximately 10 µm are the smallest grain sizes observed in xenoliths (e.g., Aupart et al., 2018; Falus et al., 2011). However, these grain sizes are likely a result of dynamic recrystallisation (e.g., Falus et al., 2011), and the "initial conditions" are not accessible. The first grains of olivine to nucleate, either during downwelling (e.g., Plümper et al., 2017) or upwelling (e.g., Satsukawa et al., 2015), are likely nm- to 100s of µm-sized. However, when the transformation to olivine is complete (e.g. from dehydration of serpentine at downwelling or from transition from wadsleyite at upwelling), the olivine grains grow at a fast rate due to the decrease in free energy (e.g. Koizumi et al., 2010; Tasaka and Hiraga, 2013). Figure S.2.9 illustrates the effect of different initial grain sizes from 10 µm to 1mm. The residence time at a certain depth or annealing time is also uncertain. Using our grain growth equation, 1 Ma to 10 Ma are necessary to provide mm-sized grains as seen in some xenoliths (e.g., Boullier and Nicolas, 1975). Furthermore, a rough estimation of a slab moving at 5 cm/year with a dip angle of 35º, would need at least a similar duration, between 5 and 10 Ma, to move from the middle of the upper mantle to the transition zone. We thus base our subsequent calculations of the grain size evolution of olivine considering residence times of 1 and 10 Ma.

Simple extrapolation of our data to geological time scales considering our best-fit parameters would indicate that olivine grain sizes (based on grain growth only) at 7 GPa and 1673 K are expected to reach 0.4 mm in approximately 1 Ma and 0.8 mm in 10 Ma. Combining these results with existing olivine flow laws (low-T plasticity: Goetze et al. (1978); dislocation creep (dry olivine): Hirth and Kohlstedt (2003), diffusion creep (dry olivine): Hirth and Kohlstedt (2003) revised by Hansen et al. (2011); disGBS (dry olivine): Hansen et al. (2011)), and considering that shear stresses are on the order of 0.1-1 MPa at a depth of ~210 km (P ≈ 7 GPa) (Kohlstedt and Hansen, 2015), olivine is expected to deform in a grain-size sensitive rheology, with similar strain rates for diffusion creep and dislocation-accommodated grain-boundary sliding (disGBS) fields (Figure S.2.10). Therefore, the results reported here would indicate that a change to a Newtonian rheology at deeper parts of the upper mantle (e.g. Karato and Wu, 1993) may be influenced by smaller olivine grain sizes due to a decrease in its growth rate. Nakakoji and Hiraga (2018) proposed that grain-boundary diffusion is a common mechanism for grain growth and diffusion creep of olivine. In short, the strain rate in the diffusion creep regime is then controlled by grain-boundary diffusion (Coble creep) and proportional to the grain-boundary diffusivity. Acknowledging the reduced diffusivities at higher pressure (e.g., Fei et al., 2016; Hirth and Kohlstedt, 2003), we anticipate that also the viscosities in the upper mantle increase.

The impact of an activation volume for grain growth on the deep upper-mantle viscosity is illustrated in Figure 8. The parameters used for the viscosity estimations are detailed in S.1. Considering the evolution of temperature with depth in the deep upper mantle (Figure 8a) and an activation volume of 0 $m^3$/mol (i.e., no pressure effect on grain growth) the grain growth rate would increase with depth, due to the increase of temperature with depth (Figure 8b). In opposition, considering an activation volume of 4.8 x $10^{-6}$ $m^3$/mol, as found here, grain sizes would only change marginally at depth. Hence, viscosities are not expected to change significantly due to pressure in the 200-400 km depth interval (Figure 8c). The expected viscosities when considering an activation volume of 4.8 x $10^{-6}$ $m^3$/mol are approximately one order of magnitude lower than when no activation volume is considered. These results suggest that mantle viscosities may be lower at increasing pressures than previously expected. This interpretation is corroborated by estimations of viscosities in the upper mantle based on postglacial rebound observations (e.g. Paulson et al., 2007).

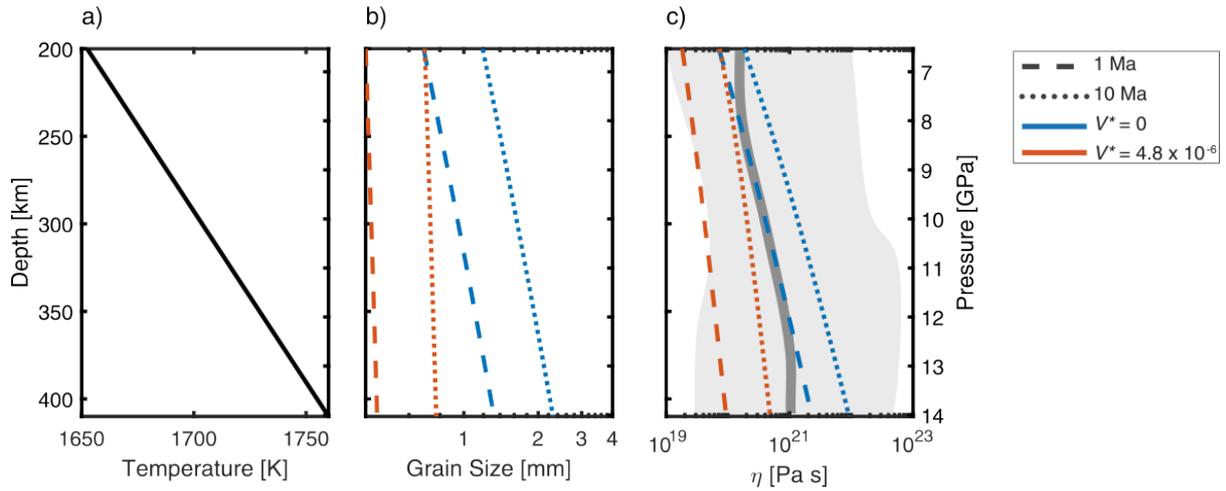

Figure 8: Viscosity estimation at deep upper mantle conditions: a) Average temperature profile of the deep upper mantle (Katsura et al., 2004). b) Expected grain sizes at 1 Ma (dashed lines) and 10 Ma (dotted lines) at the conditions of the geotherm shown in figure a, considering an initial grain size $d_0$= 10 μm, an activation volume of $V^* = 0$ m$^3$/mol (blue lines) and $V^* = 4.8 \times 10^{-6}$ m$^3$/mol (orange lines). c) Resulting viscosity profile of the deep upper mantle considering the grain size evolution shown in b, for a constant shear stress of 1 MPa and strain rates from experimental flow laws of diffusion creep (Hansen et al., 2011; Hirth and Kohlstedt, 2003), disGBS (Hansen et al., 2011), and dislocation creep (Hirth and Kohlstedt, 2003). Gray curve and shaded region illustrates a model of global average viscosity and 3-D lateral variations, respectively (Paulson et al., 2007).

## Conclusions

We investigated grain growth of olivine in dry, melt-free aggregates of olivine plus 6 and 13 vol.% of pyroxene (dunite and harzburgite, respectively), at pressure and temperature conditions of the Earth's upper mantle. Olivine is the main phase in the upper mantle and its grain size plays a major role in controlling viscosity in this region. Here we show that the olivine grain growth rate decreases as pressure increases. We propose that this effect is a result of slower grain-boundary diffusion at high pressures (Fei et al., 2016). The implications for this finding are two-fold: i) changes in grain size due to slower grain growth might influence a transition to a Newtonian rheology in deeper parts of the upper mantle, and ii) because grain-boundary diffusion is also proposed to be also the main mechanism for diffusion creep in olivine (Tasaka and Hiraga, 2013), an increase in viscosity in the deep upper mantle as a function of pressure is expected, if diffusion creep is the dominant deformation mechanism in this region, as proposed by Hirth and Kohlstedt (2003). We show that for increasing depths in the upper mantle, the increase in the grain growth rate of olivine due to an increase in temperature is counteracted by the inhibition of grain growth due to an increase in pressure. Estimating a viscosity profile as a function of depth based on the grain-size evolution of olivine predicted here and existing flow laws indicates that viscosity in the deep upper mantle is lower than previously expected.

# Acknowledgments

We thank H. Keppler for help with the synthesis of sol-gel olivine. We thank L. Tajčmanová and H.P Meyer for their support in EMP analyses. This work was supported by the Deutsche Forschungsgemeinschaft (DFG), grant no. INST 91/315-1 FUGG. KM acknowledge financial support through DFG grant MA 6287/6. MT was supported by the Visitor's program of the Bavarian Geoinstitute. SK acknowledges the support of the Earthquake Research Institute's cooperative research program to Takehiko Hiraga and by the JSPS through Grant-in-Aid for Scientific Research no. 18K03799 to SK.

# Open Research

The data files used in this paper are available at (Ferreira et al, 2021).

# References

Atkinson, A., 1985. Grain boundary diffusion - structural effects and mechanisms. J. Phys. Colloques 46, C4-379-C4-391. https://doi.org/10.1051/jphyscol:1985441

Atkinson, H.V., 1988. Overview no. 65: Theories of normal grain growth in pure single phase systems. Acta Metallurgica 36, 469–491. https://doi.org/10.1016/0001-6160(88)90079-X

Aupart, C., Dunkel, K.G., Angheluta, L., Austrheim, H., Ildefonse, B., Malthe-Sørenssen, A., Jamtveit, B., 2018. Olivine Grain Size Distributions in Faults and Shear Zones: Evidence for Nonsteady State Deformation. Journal of Geophysical Research: Solid Earth 123, 7421–7443. https://doi.org/10.1029/2018JB015836

Bäurer, M., Syha, M., Weygand, D., 2013. Combined experimental and numerical study on the effective grain growth dynamics in highly anisotropic systems: Application to barium titanate. Acta Materialia 61, 5664–5673. https://doi.org/10.1016/j.actamat.2013.06.007

Bhattacharya, A., et al., Grain boundary velocity and curvature are not correlated in Ni polycrystals. Science, 2021. 374(6564): p. 189-193

Boullier, A.M., Nicolas, A., 1975. Classification of textures and fabrics of peridotite xenoliths from South African kimberlites. Physics and Chemistry of the Earth 9, 467IN7469-468475.

Burke, J.E., Turnbull, D., 1952. Recrystallization and grain growth. Progress in Metal Physics 3, 220–292. https://doi.org/10.1016/0502-8205(52)90009-9

Cantwell, P. R., Tang, M., Dillon, S. J., Luo, J., Rohrer, G. S., & Harmer, M. P. (2014). Grain boundary complexions. Acta Materialia, 62, 1-48. https://doi.org/10.1016/j.actamat.2013.07.037

Dannberg, J., Eilon, Z., Faul, U., Gassmöller, R., Moulik, P., Myhill, R., 2017. The importance of grain size to mantle dynamics and seismological observations. Geochemistry, Geophysics, Geosystems 18, 3034–3061. https://doi.org/10.1002/2017GC006944

Davis, B.T.C., England, J.L., 1964. The melting of forsterite up to 50 kilobars. J. Geophys. Res. 69, 1113–1116. https://doi.org/10.1029/JZ069i006p01113

Edgar, A.D., 1973. Experimental petrology: basic principles and techniques. Oxford University Press.

Evans, B., Renner, J., Hirth, G., 2001. A few remarks on the kinetics of static grain growth in rocks. Int J Earth Sci 90, 88–103. https://doi.org/10.1007/s005310000150

Ezad, I.S., 2019. The kinetics of garnet breakdown and spinel grain growth in the upper mantle (PhD Thesis). UCL (University College London).

Falus, G., Tommasi, A., Soustelle, V., 2011. The effect of dynamic recrystallization on olivine crystal preferred orientations in mantle xenoliths deformed under varied stress conditions. Journal of Structural Geology 33, 1528–1540. https://doi.org/10.1016/j.jsg.2011.09.010

Farver, J.R., Yund, R.A., 2000. Silicon diffusion in forsterite aggregates: Implications for diffusion accommodated creep. Geophys. Res. Lett. 27, 2337–2340. https://doi.org/10.1029/2000GL008492

Faul, U.H., Jackson, I., 2007. Diffusion creep of dry, melt-free olivine. Journal of Geophysical Research: Solid Earth 112. https://doi.org/10.1029/2006JB004586

Faul, U.H., Scott, D., 2006. Grain growth in partially molten olivine aggregates. Contributions to Mineralogy and Petrology 151, 101–111. https://doi.org/10.1007/s00410-005-0048-1

Fei, H., Koizumi, S., Sakamoto, N., Hashiguchi, M., Yurimoto, H., Marquardt, K., Miyajima, N., Yamazaki, D., Katsura, T., 2016. New constraints on upper mantle creep mechanism inferred from silicon grain-boundary diffusion rates. Earth and Planetary Science Letters 433, 350–359. https://doi.org/10.1016/j.epsl.2015.11.014

Ferreira, F., Hansen, L.N., Marquardt, K., 2021. The Effect of Grain Boundaries on Plastic Deformation of Olivine. Journal of Geophysical Research: Solid Earth 126, e2020JB020273. https://doi.org/10.1029/2020JB020273

Furstoss, J., Hirel, P., Carrez, P., & Cordier, P. (2022). Complexions and stoichiometry of the 60.8°//[100](011) symmetrical tilt grain boundary in Mg2SiO4 forsterite: A combined empirical potential and first-principles study. American Mineralogist, 107(11), 2034-2043. https://doi.org/10.2138/am-2022-8420

Goetze, C., Poirier, J.P., Kelly, A., Cook, A.H., Greenwood, G.W., 1978. The mechanisms of creep in olivine. Philosophical Transactions of the Royal Society of London. Series A, Mathematical and Physical Sciences 288, 99–119. https://doi.org/10.1098/rsta.1978.0008

Gottstein, G., Shvindlerman, L.S., 2009. Grain boundary migration in metals: thermodynamics, kinetics, applications. CRC press.

Hall, C.E., Parmentier, E.M., 2003. Influence of grain size evolution on convective instability. Geochem. Geophys. Geosyst. 4. https://doi.org/10.1029/2002GC000308

Hansen, L.N., Zimmerman, M.E., Kohlstedt, D.L., 2011. Grain boundary sliding in San Carlos olivine: Flow law parameters and crystallographic-preferred orientation. Journal of Geophysical Research 116. https://doi.org/10.1029/2011JB008220

Heilbronner, R., Bruhn, D., 1998. The influence of three-dimensional grain size distributions on the rheology of polyphase rocks. Journal of Structural Geology 20, 695–705. https://doi.org/10.1016/S0191-8141(98)00010-8

Heinemann, S., Wirth, R., Gottschalk, M., Dresen, G., 2005. Synthetic [100] tilt grain boundaries in forsterite: 9.9 to 21.5°. Phys Chem Minerals 32, 229–240. https://doi.org/10.1007/s00269-005-0448-9

Hench, L.L., West, J.K., 1990. The sol-gel process. Chem. Rev. 90, 33–72. https://doi.org/10.1021/cr00099a003

Hielscher, R., Schaeben, H., 2008. A novel pole figure inversion method: specification of the MTEX algorithm. Journal of Applied Crystallography 41, 1024–1037.

Hillert, M., 1965. On the theory of normal and abnormal grain growth. Acta Metallurgica 13, 227–238. https://doi.org/10.1016/0001-6160(65)90200-2

Hiraga, T., Tachibana, C., Ohashi, N., Sano, S., 2010. Grain growth systematics for forsterite ± enstatite aggregates: Effect of lithology on grain size in the upper mantle. Earth and Planetary Science Letters 291, 10–20. https://doi.org/10.1016/j.epsl.2009.12.026

Hirth, G., Kohlstedt, D., 2003. Rheology of the upper mantle and the mantle wedge: A view from the experimentalists, in: Eiler, J. (Ed.), Geophysical Monograph Series. American Geophysical Union, Washington, D. C., pp. 83–105.

Humphreys, F.J., Hatherly, M., 2004. Recrystallization and related annealing phenomena. Elsevier, Amsterdam; Boston.

Jackson, I., Fitz Gerald, J.D., Faul, U.H., Tan, B.H., 2002. Grain-size-sensitive seismic wave attenuation in polycrystalline olivine: Seismic wave attenuation in olivine. Journal of Geophysical Research: Solid Earth 107, ECV 5-1-ECV 5-16. https://doi.org/10.1029/2001JB001225

Karato, S., 2008. Deformation of Earth Materials: An Introduction to the Rheology of Solid Earth. Cambridge University Press, Cambridge. https://doi.org/10.1017/CBO9780511804892

Karato, S., 1992. On the Lehmann discontinuity. Geophysical Research Letters 19, 2255–2258.

Karato, S., 1989. Grain growth kinetics in olivine aggregates. Tectonophysics 168, 255–273. https://doi.org/10.1016/0040-1951(89)90221-7

Karato, S. -i., Wu, P., 1993. Rheology of the Upper Mantle: A Synthesis. Science 260, 771–778. https://doi.org/10.1126/science.260.5109.771

Katsura, T., Yamada, H., Nishikawa, O., Song, M., Kubo, A., Shinmei, T., Yokoshi, S., Aizawa, Y., Yoshino, T., Walter, M.J., Ito, E., Funakoshi, K., 2004. Olivine-wadsleyite transition in the system (Mg,Fe)2SiO4. Journal of Geophysical Research: Solid Earth 109. https://doi.org/10.1029/2003JB002438

Kelly, M. N., Rheinheimer, W., Hoffmann, M. J., & Rohrer, G. S. (2018). Anti-thermal grain growth in SrTiO: Coupled reduction of the grain boundary energy and grain growth rate constant. Acta Materialia, *149*, 11-18. https://doi.org/10.1016/j.actamat.2018.02.030

Kohlstedt, D.L., Hansen, L.N., 2015. Constitutive Equations, Rheological Behavior, and Viscosity of Rocks, in: Treatise on Geophysics. Elsevier, pp. 441–472. https://doi.org/10.1016/B978-0-444-53802-4.00042-7

Koizumi, S., Hiraga, T., Tachibana, C., Tasaka, M., Miyazaki, T., Kobayashi, T., Takamasa, A., Ohashi, N., Sano, S., 2010. Synthesis of highly dense and fine-grained aggregates of mantle composites by vacuum sintering of nano-sized mineral powders. Physics and Chemistry of Minerals 37, 505–518. https://doi.org/10.1007/s00269-009-0350-y

Lehmann, I., 1961. S and the structure of the upper mantle. Geophyical Journal of the Royal Astronomical Society 124–138.

Lehmann, I., 1959. Velocities of longitudinal waves in the upper part of the earth's mantle. Annales de Geophysique 15, 93.

Liebermann, R.C., Wang, Y., 2013. Characterization of Sample Environment in a Uniaxial Split-Sphere Apparatus, in: Syono, Y., Manghnani, M.H. (Eds.), Geophysical Monograph Series. American Geophysical Union, Washington, D. C., pp. 19–31. https://doi.org/10.1029/GM067p0019

Marquardt, K., Faul, U.H., 2018. The structure and composition of olivine grain boundaries: 40 years of studies, status and current developments. Phys Chem Minerals 45, 139–172. https://doi.org/10.1007/s00269-017-0935-9

Marquardt, K., Graef, M.D., Singh, S., Marquardt, H., Rosenthal, A., Koizuimi, S., 2017. Quantitative electron backscatter diffraction (EBSD) data analyses using the dictionary indexing (DI) approach: Overcoming indexing difficulties on geological materials. American Mineralogist 102, 1843–1855. https://doi.org/10.2138/am-2017-6062

Marquardt, K., Rohrer, G.S., Morales, L., Rybacki, E., Marquardt, H., Lin, B., 2015. The most frequent interfaces in olivine aggregates: the GBCD and its importance for grain boundary related processes. Contributions to Mineralogy and Petrology 170. https://doi.org/10.1007/s00410-015-1193-9

Mendelson, M.I., 1969. Average Grain Size in Polycrystalline Ceramics. Journal of the American Ceramic Society 52, 443–446. https://doi.org/10.1111/j.1151-2916.1969.tb11975.x

Mulyukova, E., Bercovici, D., 2019. The Generation of Plate Tectonics From Grains to Global Scales: A Brief Review. Tectonics 38, 4058–4076. https://doi.org/10.1029/2018TC005447

Nakakoji, T., Hiraga, T., 2018. Diffusion Creep and Grain Growth in Forsterite +20 vol% Enstatite Aggregates: 2. Their Common Diffusional Mechanism and Its Consequence for Weak-Temperature-Dependent Viscosity. Journal of Geophysical Research: Solid Earth 123, 9513–9527. https://doi.org/10.1029/2018JB015819

Nes, E., Ryum, N., Hunderi, O., 1985. On the Zener drag. Acta Metallurgica 33, 11–22. https://doi.org/10.1016/0001-6160(85)90214-7

Nichols, S.J., Mackwell, S.J., 1991. Grain growth in porous olivine aggregates. Phys Chem Minerals 18, 269–278. https://doi.org/10.1007/BF00202580

Nishihara, Y., Doi, S., Kakizawa, S., Higo, Y., Tange, Y., 2020. Effect of pressure on temperature measurements using WRe thermocouple and its geophysical impact. Physics of the Earth and Planetary Interiors 298, 106348. https://doi.org/10.1016/j.pepi.2019.106348

Ohashi, Y., 1984. Polysynthetically-twinned structures of enstatite and wollastonite. Phys Chem Minerals 10, 217–229. https://doi.org/10.1007/BF00309314

Ohtani, E., Kumazawa, M., 1981. Melting of forsterite Mg2SiO4 up to 15 GPa. Physics of the Earth and Planetary Interiors 27, 32–38. https://doi.org/10.1016/0031-9201(81)90084-4

Ohuchi, T., Nakamura, M., 2007a. Grain growth in the forsterite–diopside system. Physics of the Earth and Planetary Interiors 160, 1–21. https://doi.org/10.1016/j.pepi.2006.08.003

Ohuchi, T., Nakamura, M., 2007b. Grain growth in the system forsterite–diopside–water. Physics of the Earth and Planetary Interiors 161, 281–304. https://doi.org/10.1016/j.pepi.2007.02.009

Paulson, A., Zhong, S., Wahr, J., 2007. Limitations on the inversion for mantle viscosity from postglacial rebound. Geophysical Journal International 168, 1195–1209. https://doi.org/10.1111/j.1365-246X.2006.03222.x

Plümper, O., John, T., Podladchikov, Y.Y., Vrijmoed, J.C., Scambelluri, M., 2017. Fluid escape from subduction zones controlled by channel-forming reactive porosity. Nature Geosci 10, 150–156. https://doi.org/10.1038/ngeo2865

Pommier, A., Kohlstedt, D.L., Hansen, L.N., Mackwell, S., Tasaka, M., Heidelbach, F., Leinenweber, K., 2018. Transport properties of olivine grain boundaries from electrical conductivity experiments. Contributions to Mineralogy and Petrology 173. https://doi.org/10.1007/s00410-018-1468-z

Rheinheimer, W., Bäurer, M., Chien, H., Rohrer, G.S., Handwerker, C.A., Blendell, J.E., Hoffmann, M.J., 2015. The equilibrium crystal shape of strontium titanate and its relationship to the grain boundary plane distribution. Acta Materialia 82, 32–40. https://doi.org/10.1016/j.actamat.2014.08.065

Ringwood, A.E., 1970. Phase transformations and the constitution of the mantle. Physics of the Earth and Planetary Interiors 3, 109–155. https://doi.org/10.1016/0031-9201(70)90047-6

Rollett, A.D., Mullins, W.W., 1997. On the growth of abnormal grains. Scripta Materialia 36, 975–980. https://doi.org/10.1016/S1359-6462(96)00501-5

Rozel, A., 2012. Impact of grain size on the convection of terrestrial planets: CONVECTION AND GRAIN SIZE. Geochem. Geophys. Geosyst. 13, n/a-n/a. https://doi.org/10.1029/2012GC004282

Rubie, D.C., Karato, S., Yan, H., O'Neill, H.St.C., 1993. Low differential stress and controlled chemical environment in multianvil high-pressure experiments. Phys Chem Minerals 20, 315–322. https://doi.org/10.1007/BF00215102

Satsukawa, T., Griffin, W.L., Piazolo, S., O'Reilly, S.Y., 2015. Messengers from the deep: Fossil wadsleyite-chromite microstructures from the Mantle Transition Zone. Scientific Reports 5, 16484. https://doi.org/10.1038/srep16484

Smith, C.S., 1948. Grains, phases, and interfaces: An introduction of microstructure. Trans. Metall. Soc. AIME 175, 15–51.

Solomatov, V.S., 2001. Grain size-dependent viscosity convection and the thermal evolution of the Earth. Earth and Planetary Science Letters 191, 203–212. https://doi.org/10.1016/S0012-821X(01)00426-5

Solomatov, V.S., El-Khozondar, R., Tikare, V., 2002. Grain size in the lower mantle: constraints from numerical modeling of grain growth in two-phase systems. Physics of the Earth and Planetary Interiors 129, 265–282. https://doi.org/10.1016/S0031-9201(01)00295-3

Solomatov, V.S., Reese, C.C., 2008. Grain size variations in the Earth's mantle and the evolution of primordial chemical heterogeneities. J. Geophys. Res. 113, B07408. https://doi.org/10.1029/2007JB005319

Speciale, P.A., Behr, W.M., Hirth, G., Tokle, L., 2020. Rates of Olivine Grain Growth During Dynamic Recrystallization and Postdeformation Annealing. Journal of Geophysical Research: Solid Earth 125, e2020JB020415. https://doi.org/10.1029/2020JB020415

Tan, B.H., Jackson, I., Fitz Gerald, J.D., 1997. Shear wave dispersion and attenuation in fine-grained synthetic olivine aggregates: Preliminary results. Geophys. Res. Lett. 24, 1055–1058. https://doi.org/10.1029/97GL00860

Tasaka, M., Hiraga, T., 2013. Influence of mineral fraction on the rheological properties of forsterite + enstatite during grain-size-sensitive creep: 1. Grain size and grain growth laws. Journal of Geophysical Research: Solid Earth 118, 3970–3990. https://doi.org/10.1002/jgrb.50285

ten Grotenhuis, S.M., Drury, M.R., Peach, C.J., Spiers, C.J., 2004. Electrical properties of fine-grained olivine: Evidence for grain boundary transport. Journal of Geophysical Research: Solid Earth 109. https://doi.org/10.1029/2003JB002799

Thielmann, M., 2018. Grain size assisted thermal runaway as a nucleation mechanism for continental mantle earthquakes: Impact of complex rheologies. Tectonophysics 746, 611–623. https://doi.org/10.1016/j.tecto.2017.08.038

Thielmann, M., Rozel, A., Kaus, B.J.P., Ricard, Y., 2015. Intermediate-depth earthquake generation and shear zone formation caused by grain size reduction and shear heating. Geology 43, 791–794. https://doi.org/10.1130/G36864.1

van Driel, J., Schusteritsch, G., Brodholt, J.P., Dobson, D.P., Pickard, C.J., 2020. The discontinuous effect of pressure on twin boundary strength in MgO. Phys Chem Minerals 47, 11. https://doi.org/10.1007/s00269-019-01079-1

Walter, M.J., Thibault, Y., Wei, K., Luth, R.W., 1995. Characterizing experimental pressure and temperature conditions in multi-anvil apparatus. Can. J. Phys. 73, 273–286. https://doi.org/10.1139/p95-039

Watson, E., Wark, D., Price, J., Van Orman, J., 2002. Mapping the thermal structure of solid-media pressure assemblies. Contrib Mineral Petrol 142, 640–652. https://doi.org/10.1007/s00410-001-0327-4

Wright, S.I., Nowell, M.M., Lindeman, S.P., Camus, P.P., De Graef, M., Jackson, M.A., 2015. Introduction and comparison of new EBSD post-processing methodologies. Ultramicroscopy 159, 81–94. https://doi.org/10.1016/j.ultramic.2015.08.001

Yokoi, T., Yoshiya, M., 2018. Atomistic simulations of grain boundary transformation under high pressures in MgO. Physica B: Condensed Matter, Special issue on Frontiers in Materials Science: Condensed Matters 532, 2–8. https://doi.org/10.1016/j.physb.2017.03.014

Zhang, Z., Karato, S.-I., 2021. The Effect of Pressure on Grain growth Kinetics in Olivine Aggregates With Some Geophysical Applications. Journal of Geophysical Research: Solid Earth 126, e2020JB020886. https://doi.org/10.1029/2020JB020886

Filippe Ferreira[1], Marcel Thielmann[1], Robert Farla[2], Sanae Koizumi[3], Katharina Marquardt[1,4*]

1 Bayerisches Geoinstitut, Universität Bayreuth, Bayreuth, Germany
2 Deutsches Elektronen-Synchrotron (DESY), Hamburg, Germany
3 Earthquake Research Institute, Tokyo, Japan
4 Department of Materials. University of Oxford, Oxford, UK

*Corresponding author: Katharina Marquardt (Katharina.marquardt@materials.ox.ac.uk)

**Contents of this file**

# Supplemental material
## S.1: Calculation of viscosity profiles

Viscosity, $\eta$, as shown in the viscosity profiles in Figure 9, is given by the ratio between the shear stress, $\sigma$, and strain rate, $\dot{\varepsilon}$, that is:

$$\eta = \frac{\sigma}{\dot{\varepsilon}} \quad \text{(E1)}$$

Our calculations assume $\sigma = 1$ MPa (Kohlstedt and Hansen, 2015), and the strain rate is given by the sum of the strain rates of dislocation creep ($\dot{\varepsilon}_{dis}$), dislocation-accommodated grain-boundary sliding ($\dot{\varepsilon}_{disGBS}$) and diffusion creep ($\dot{\varepsilon}_{dif}$):

$$\dot{\varepsilon} = \dot{\varepsilon}_{dis} + \dot{\varepsilon}_{GBS} + \dot{\varepsilon}_{dif} \quad \text{(E2)}$$

The strain rate for dislocation creep is given by (dry olivine, Hirth and Kohlstedt, 2003):

$$\dot{\varepsilon}_{dis} = (1.1 \times 10^5) \exp\left(\frac{-(5.3 \times 10^5) + P(14 \times 10^{-6})}{RT}\right) \quad \text{(E3)}$$

for dislocation-accommodated grain-boundary sliding by (dry olivine; Hirth and Kohlstedt, 2003; modified by Hansen et al., 2011):

$$\dot{\varepsilon}_{GBS} = (10^{4.8}) \frac{1}{d^{0.73}} \exp\left(\frac{-(4.45 \times 10^5 - (300 \times 10^6 \times 18 \times 10^{-6})) + P(18 \times 10^{-6})}{RT}\right) \quad \text{(E4)}$$

and for diffusion creep by (dry olivine; Hirth and Kohlstedt, 2003):

$$\dot{\varepsilon}_{dif} = (10^{7.6}) \frac{1}{d^3} \exp\left(\frac{-(3.7 \times 10^5) + P(6 \times 10^{-6})}{RT}\right) \quad \text{(E5)}$$

where $P$ is pressure in GPa $R$ is the gas constant in J·K$^{-1}$·mol$^{-1}$, $T$ is the temperature in K, and $d$ is grain size in μm. The grain-size evolution as a function of pressure, $P$

(Pa), temperature, $T$ (K), and time, $t$ (s), is given by Eq. 1 and Eq. 2, using the parameters obtained in this study, that is:

$$d = \left(d_0^{3.88} + \left((2.11 \times 10^{-7}) \exp\left(\frac{-(608\times10^3)+P(4.3\times10^{-6})}{RT}\right)\right)t\right)^{1/3.88} \qquad (E6)$$

Pressure and temperature as a function of depth were obtained from Dziewonski and Anderson (1981) and Katsura et al. (2004), respectively. Our calculations were performed using an initial grain size, $d_0$, of $10 \times 10^{-6}$ m.

## S.2: Supplemental Figures

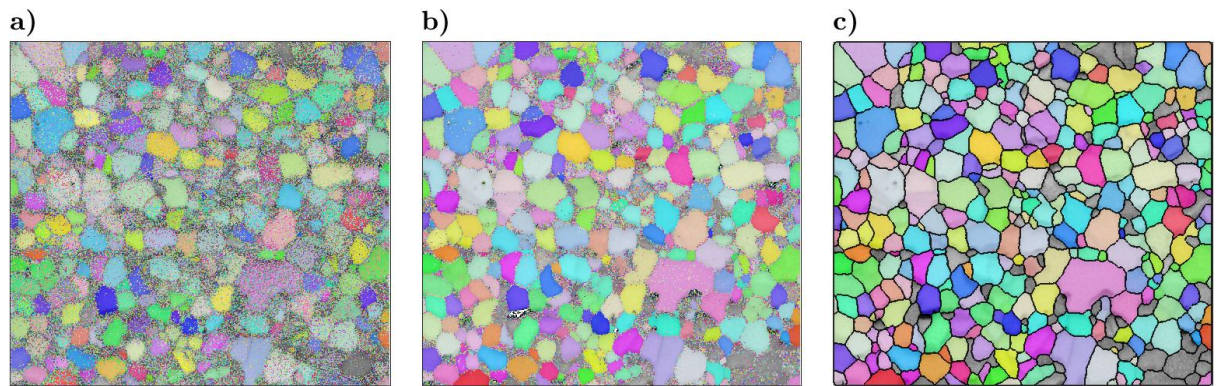

Figure S.2.1: Cleaning of EBSD data exemplified with the sample FSG5 - starting material. EBSD data is shown by inverse pole figure maps of olivine overlaying grayscale image quality maps: a) EBSD data as acquired, b) after NPAR re-indexing and c) after pseudo-symmetry correction and grain-dilation. Resulting grain boundaries are also shown in this figure. Images are 35 x 35 µm wide. Grey areas are poorly indexed or pyroxene grains.

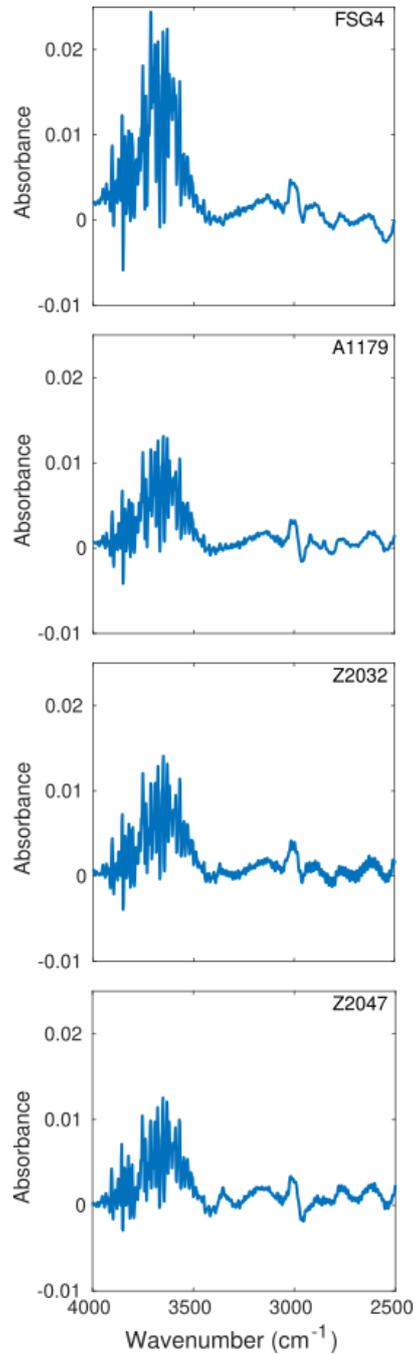

Figure S.2.2: Representative FTIR spectrum measured for the starting material after sintering (FSG4) and after annealing experiments at 1 GPa (A1179), 7GPa (Z2032) and 10 GPa (Z2047). Note the absence of sharp peaks in the 2950 cm$^{-1}$ 3750 cm$^{-1}$ range and of a broad peak around 3400 cm-1. The noise is due to atmospheric moisture adsorbed to the sample surface.

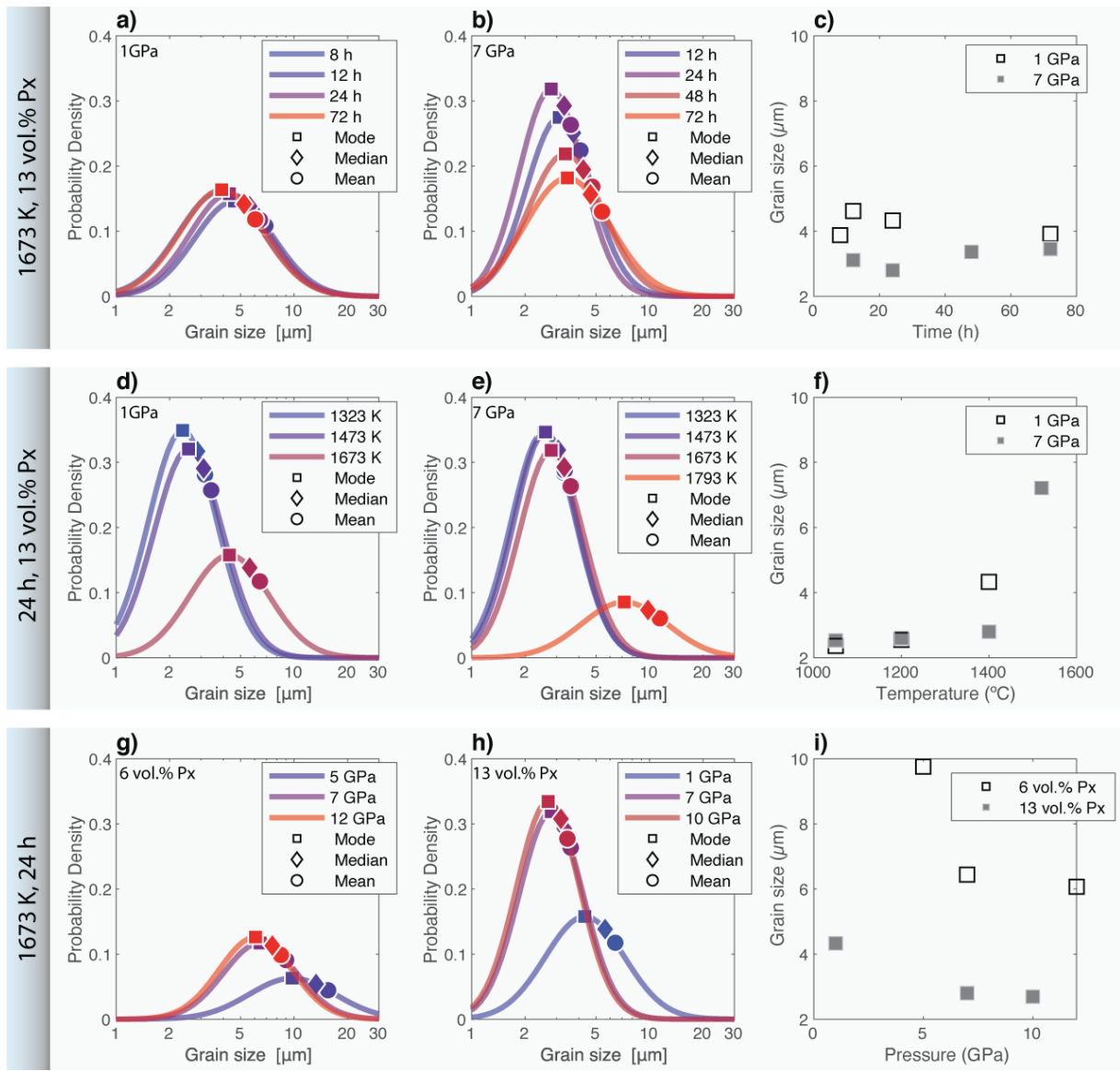

Figure S.2.3: Lognormal fit to the grain-size distributions: The upper row (a-c) shows a time series of experiments performed at 1673 K and pressures of a) 1 GPa and b) 7GPa. c) Mode of the fit to the grain-size distribution as a function of time. The middle row (d-f) shows the temperature series of experiments performed for 24 hours at pressures of d) 1 GPa and e) 7GPa. f) Mode of the of the fit to the grain-size distribution as a function of temperature. The bottom row (g-i) shows the pressure series of experiments performed at 1673 K for 24 hours for samples containing g) 6 vol.% and h) 13 vol.% of pyroxene. i) Mode of the fit to the grain-size distribution as a function of pressure for samples with a Pyroxene content of 6 vol.% and 13 vol.%.

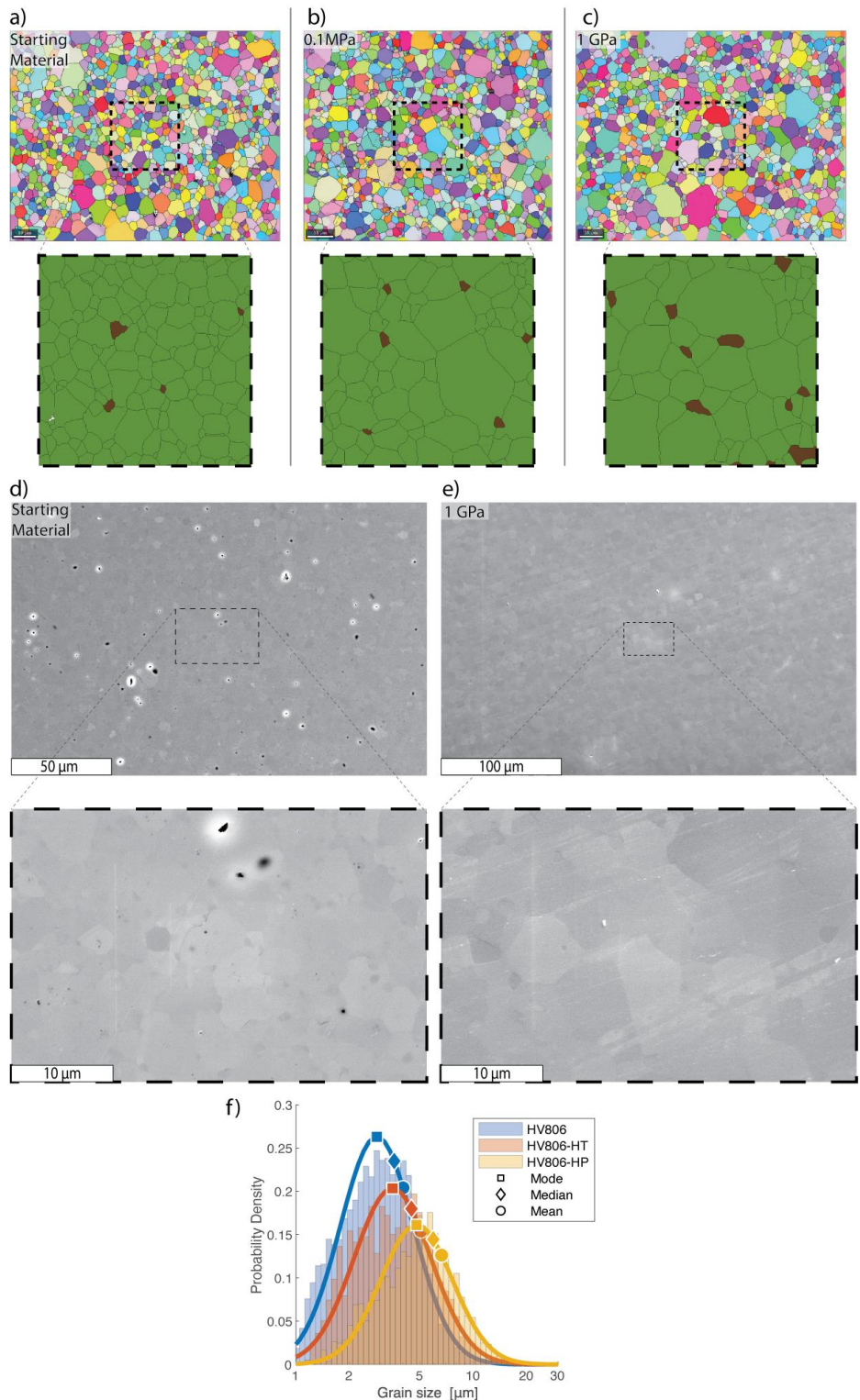

Figure S.2.4. Microstructure of samples a)HV806 (vacuum sintered Fo+10 vol.% En, starting material), b) HV806-HT (1673 K, 24 h, 0.1 MPa) and c) HV806-HP (1673 K, 24 h, 1 GPa). Images show inverse pole figure maps of olivine with insets showing phase maps: olivine is green and pyroxene is brown. Secondary electron image of samples d) HV806 and e) HV806-HP. Insets show magnified regions. Insets are 50 μm x 50μm. Note pores and holes in the starting material (d) and its virtually absence after high pressure experiments(e). f) Grain size distribution (bars) and its respective log-normal distribution (curves) of samples HV806, HV806-HT and HV806-HP. The mode, median and mean of each distribution is shown by squares, diamonds and circles respectively.

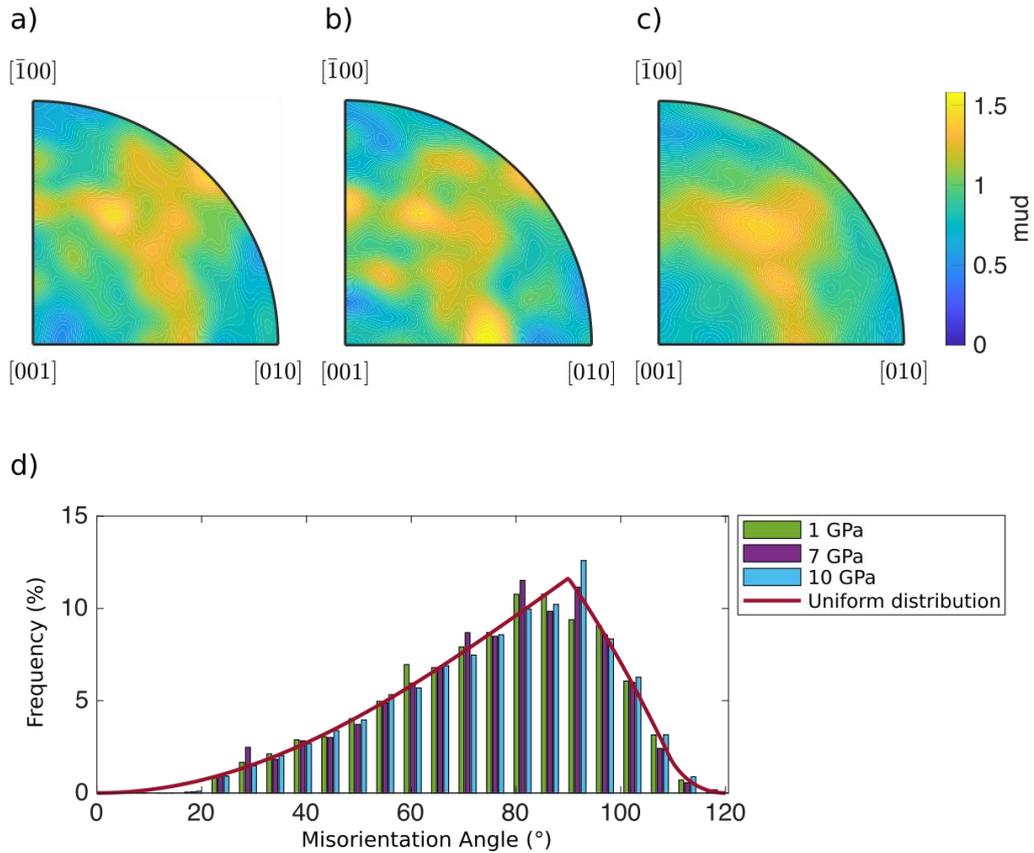

Figure S.2.5. Grain boundary misorientation analyses: Upper row: Misorientation axis distribution among olivine-olivine grain boundaries for recovered samples containing 13vol.% Px, for experiments performed at 1673 K, 24h at a) 1 GPa, b) 7 GPa and c) 10 GPa. d) Histogram of the correlated misorientation-angle distribution for the same samples. Curve shows expected uniform distribution of misorientation angles for olivine.

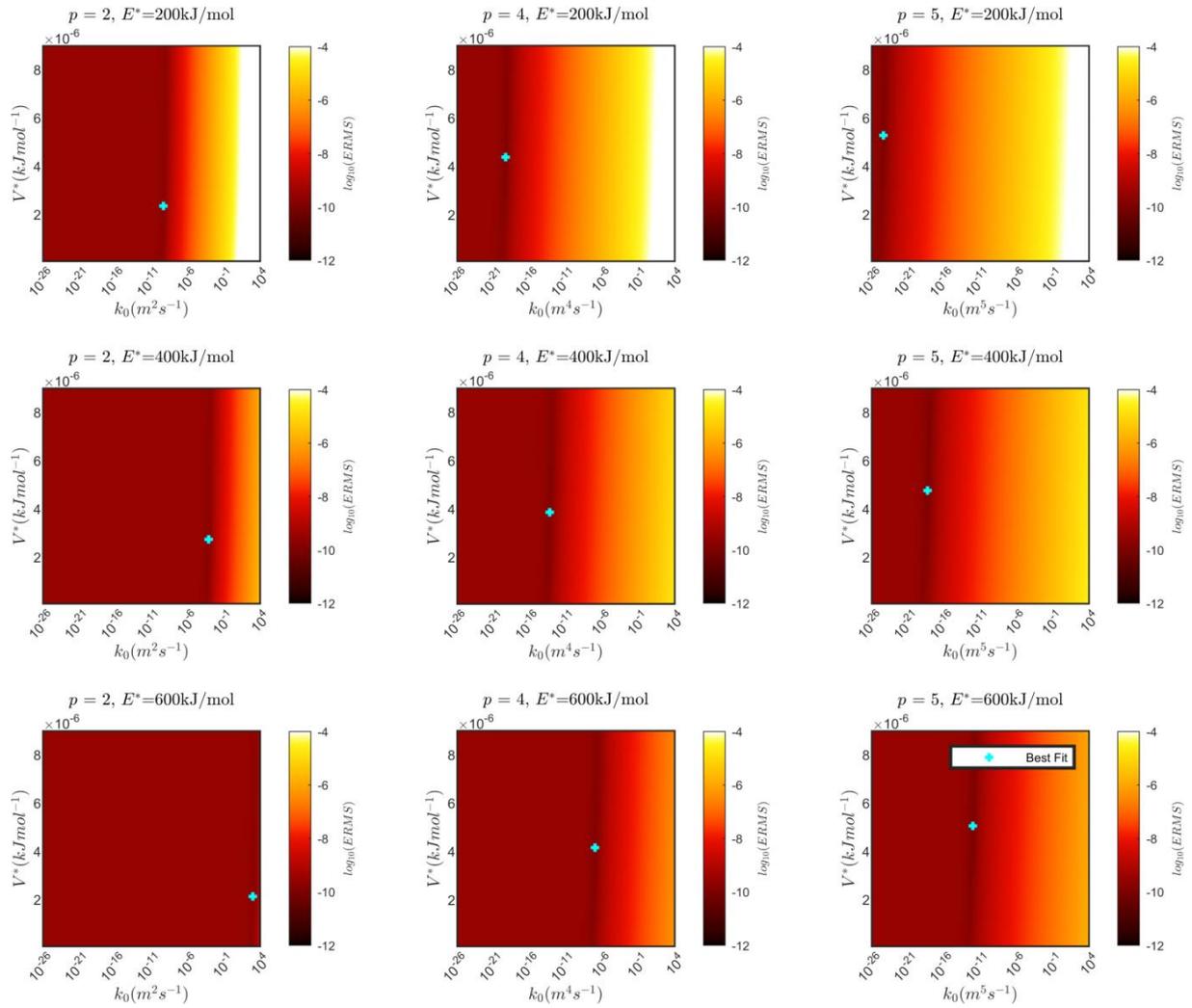

Figure S.2.6. Root-mean-squared error (*RMSE*) as a function of $k_0$ and $V^*$ for different combinations of fixed activation energy $E^*$ and grain growth exponent *p*. Blue cross indicates the minimum RMSE for each figure. All plots are obtained with temperature *T*=1673 K.

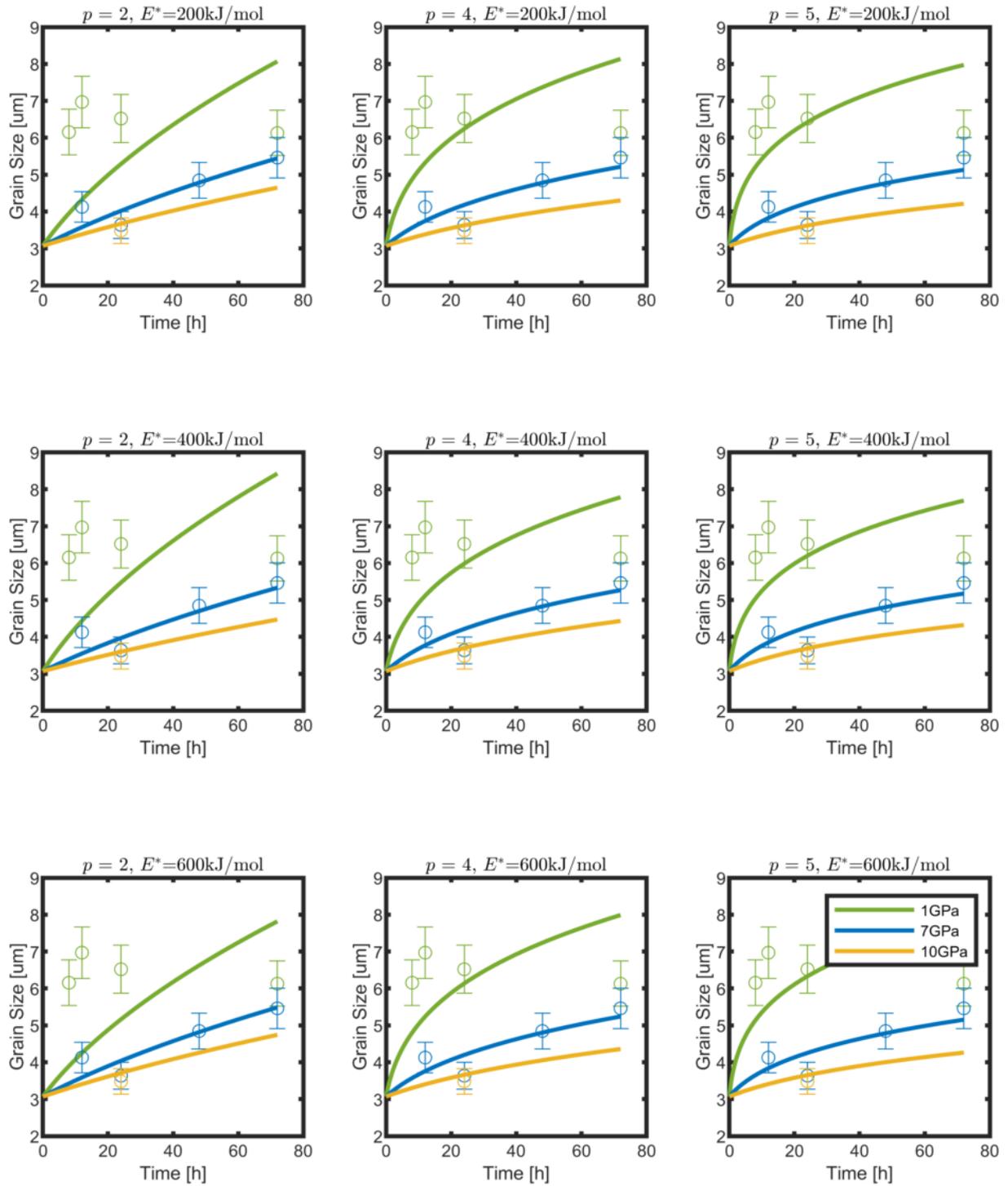

Figure S.2.7. Comparison between experimentally-obtained grain size (13 vol. % *Px, T*=1673 K) and predicted grain sizes (curves) for different combinations of fixed activation energy $E^*$ and grain growth exponent *p*.

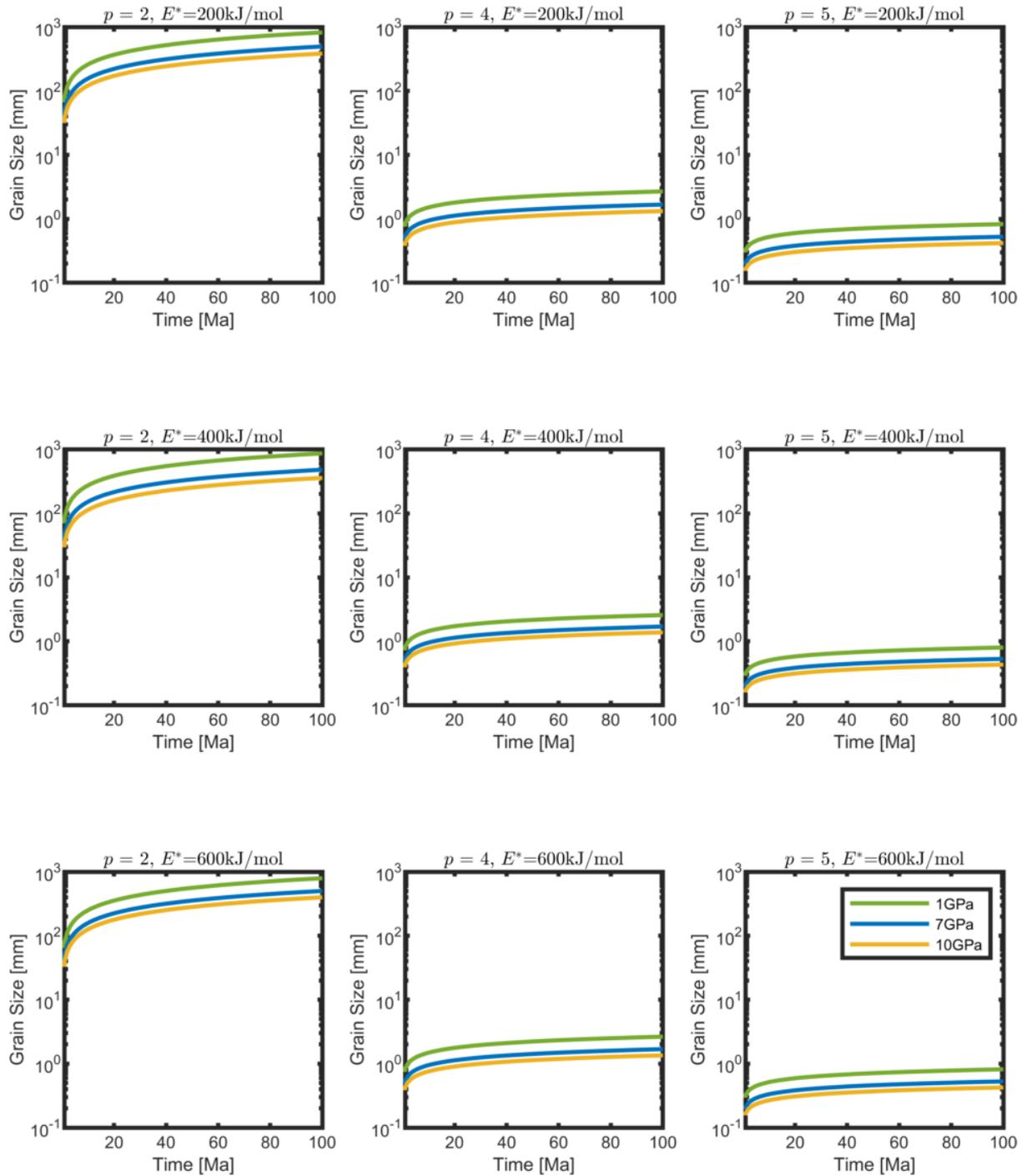

Figure S.2.8. Extrapolation of predicted grain sizes to 100 Ma for different combinations of fixed activation energy $E^*$ and grain growth exponent $p$. All plots are obtained with temperature $T$=1673 K.

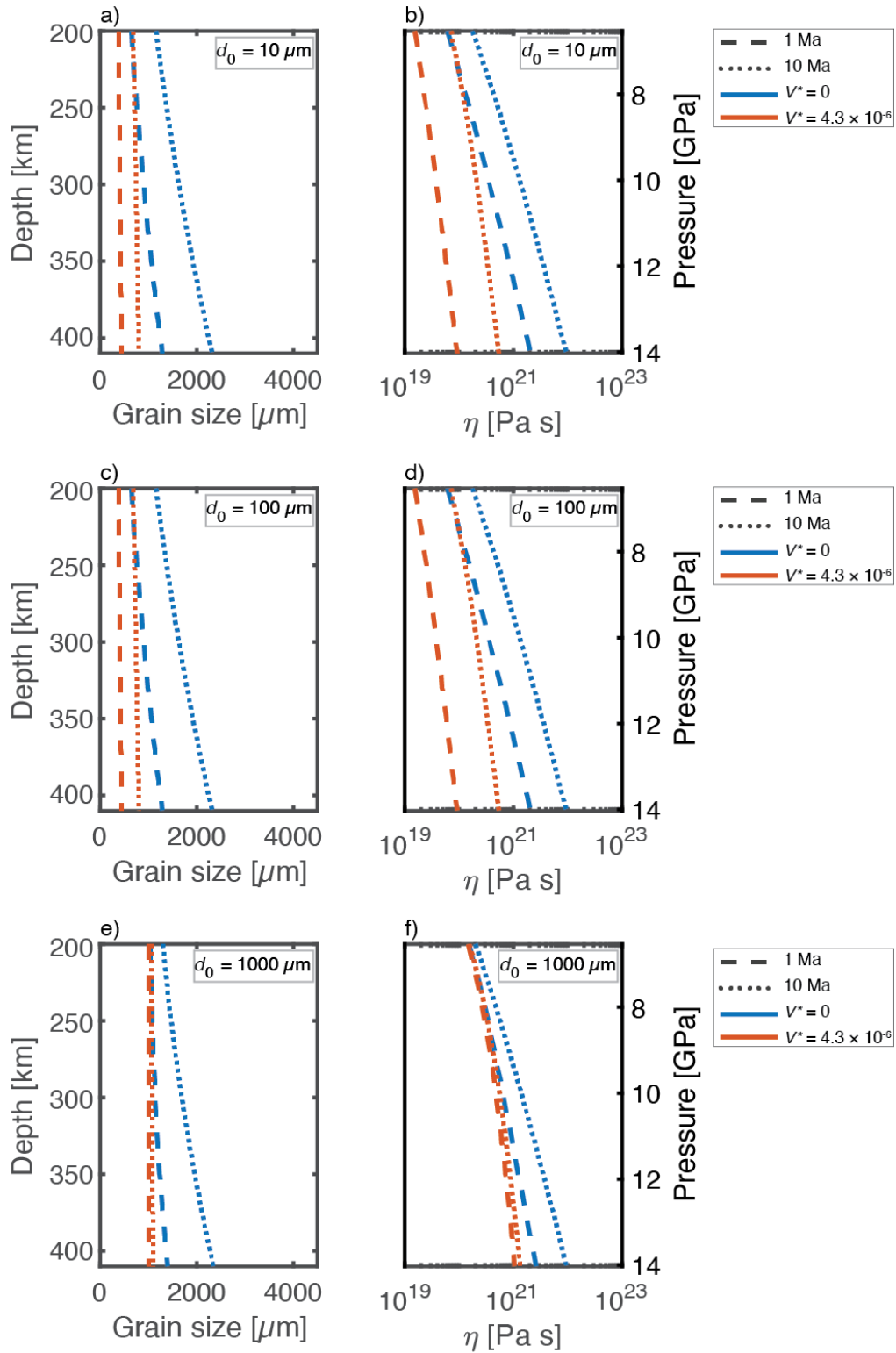

Figure S.2.9. The effect of different initial grain sizes on the viscosity estimation at deep upper mantle conditions: Figures on the left column demonstrate the expected grain sizes at 1 Ma (dashed lines) and 10 Ma (dotted lines) at the conditions of the geotherm shown in figure 9a, considering an activation volume of $V^* = 0$ m$^3$/mol (blue lines) and $V^* = 4.3 \times 10^{-6}$ m$^3$/mol (orange lines). Figures on the right column show the resulting viscosity profile of the deep upper mantle at 1 Ma and 10 Ma, for constant shear stress of 1 MPa and strain rates from experimental flow laws of disGBS, diffusion creep (Hansen et al., 2011; Hirth and Kohlstedt, 2003) and dislocation creep (Hirth and Kohlstedt, 2003). Figures in the upper(a,

b), middle(c, d) and bottom row were calculated considering initial grain sizes of 10 μm, 100 μm and 1000 μm (1 mm), respectively.

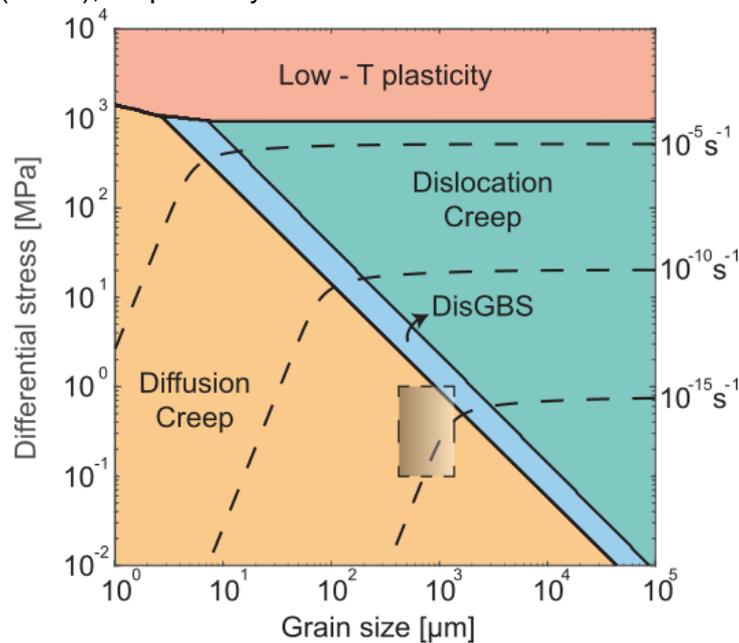

Figure S.2.10: Deformation mechanism map for olivine at 1673 K and 7 GPa. Dashed rectangle shows the region for predicted olivine grain sizes after an annealing time between 1 Ma and 100 Ma and expected differential stresses at approximately 210 km depth (7 GPa). The flow laws used in the construction of the deformation mechanism map of olivine are from: low-T plasticity: Goetze et al. (1978), dislocation creep: Hirth and Kohlstedt (2003), diffusion creep: Hirth and Kohlstedt (2003) revised by Hansen et al. (2011) and disGBS: Hansen et al. (2011).

## S.3: Supplemental Table

Table S.3.1: Electron Microprobe chemical analyses. Analyses were done in the samples Z1962 (6 vol.% Px) and Z2049 (13 vol.% Px), in spot and grid modes. The number of analyses (n) is indicated for each sample and phase.

Olivine Spot Analyses

|  | Z1962 (n=10) | | Z2049 (n=10) | |
| --- | --- | --- | --- | --- |
| Oxide | Average (wt%) | Standard Deviation | Average (wt%) | Standard Deviation |
| MgO | 49.37 | 0,11 | 49,11 | 0,28 |
| FeO | 10,32 | 0,03 | 10,71 | 0,06 |
| $SiO_2$ | 40.44 | 0,22 | 40,31 | 0,23 |

Pyroxene Spot Analyses

|  | Z1962 (n=6) | | Z2049 (n=4) | |
| --- | --- | --- | --- | --- |
| Oxide | Average (wt%) | Standard Deviation | Average (wt%) | Standard Deviation |
| MgO | 35,79 | 0,14 | 36,03 | 0,47 |
| FeO | 6,33 | 0,03 | 6,65 | 0,15 |
| $SiO_2$ | 57,29 | 0,40 | 56,65 | 0,70 |

Grid Analyses

|  | Z1962 (n=130) | | Z2049 (n=137) | |
| --- | --- | --- | --- | --- |
| Oxide | Average (wt%) | Standard Deviation | Average (wt%) | Standard Deviation |
| MgO | 47,57 | 0,69 | 46,28 | 1,02 |
| FeO | 9,60 | 0,19 | 9,64 | 0,32 |
| $SiO_2$ | 41,67 | 0,90 | 42,87 | 1,34 |

Table S.3.2: Experimental data of samples HV806, the vacuum sintered starting material, HV806-HP, the grain growth experiment at 1 GPa and HV806-HT, the grain growth experiment at 0.1 MPa (atmospheric pressure). $P$ is pressure in GPa, $T$ temperature in K, $t$ experimental duration in hours, $d$ average grain size in µm, $d_{MIL}$ the mean intercept length in µm. $Mo_{Fit}$, $\mu_{Fit}$ and $\sigma_{Fit}$ are the mode, mean and standard deviation of the lognormal fit to the grain size distribution, respectively. $f_{Px}$ is the pyroxene fraction as measured by EBSD and $n$ is the number of grains analysed for each sample.

| Sample | Starting Material | $P$ (GPa) | $T$ (K) | $t$ (h) | $d$ (µm) | $d_{MIL}$ (µm) | $Mo_{FIT}$ | $\mu_{FIT}$ | $\sigma_{FIT}$ | $f_{Px}$ (EBSD) | $n$ |
| --- | --- | --- | --- | --- | --- | --- | --- | --- | --- | --- | --- |
| HV806 | Vacuum sintered Fo + 10% Px | $10^{-11}$ | 1523 | 3 | 4.0 | 2.6 | 2.85 | 1.27 | 0.48 | 0.01 | 5424 |
| HV806-HP | HV806 | 1 | 1673 | 24 | 6.59 | 4.14 | 4.77 | 1.78 | 0.47 | 0.02 | 3013 |
| HV806-HT | HV806 | $10^{-4}$ | 1673 | 24 | 5.04 | 3.38 | 3.49 | 1.5 | 0.5 | 0.02 | 2415 |